\documentstyle[12pt,aps,psfig,eqsecnum,tighten,preprint]{revtex}
\begin{document}

\title{Haldane Gapped Spin Chains:
Exact Low Temperature Expansions of Correlation Functions}
\author{Robert M. Konik}
\address{Department of Physics, 
University of Virginia, Charlottesville, VA 22904}

\date{\today}
\maketitle
\begin{abstract}
We study both the static and dynamic properties of
gapped, one-dimensional, Heisenberg, anti-ferromagnetic, spin chains at
finite temperature through an analysis of the $O(3)$ non-linear sigma
model.  Exploiting the integrability of this theory, we are able
to compute an exact low temperature expansion of the finite
temperature correlators.  We do so using a truncated `form-factor'
expansion and so provide evidence that this technique can be successfully
extended to finite temperature.  As a direct test, we compute the static
zero-field susceptibility and obtain an exact match to the susceptibility
derived from the low temperature expansion of the exact free energy.  We also
study transport properties, computing both the spin conductance and the
NMR-relaxation rate, $1/T_1$.  We find these quantities
to show ballistic behaviour.  In particular, the computed spin conductance 
exhibits a non-zero Drude weight at finite temperature and zero applied
field.
The physics thus described
differs from the spin diffusion reported 
by Takigawa et al. \cite{takigawa} from experiments
on the Haldane gap material, $AgVP_2S_6$. 
\end{abstract}
\pacs{PACS numbers: ????}


\newcommand{\del}{\partial}
\newcommand{\nn}{\nonumber}
\newcommand{\rtc}{\tilde{\rho}_c}
\newcommand{\th}{\theta}
\newcommand{\tth}{\tilde{\theta}}
\newcommand{\gcc}{\Gamma_{cc}}
\newcommand{\la}{\lambda}
\newcommand{\CO}{{\cal O}}
\newcommand{\om}{{\omega}}
\newcommand{\ep}{{\epsilon}}
\newcommand{\ut}{{\th_{32}}}
\newcommand{\vt}{{\th_{31}}}
\newcommand{\wt}{{\th_{43}}}
\newcommand{\lb}{{\langle}}
\newcommand{\rb}{{\rangle}}
\newcommand{\dt}{{d\th \over 2\pi}}
\newcommand{\dto}{{d\th_1 \over 2\pi}}
\newcommand{\dtt}{{d\th_2 \over 2\pi}}
\newcommand{\dttr}{{d\th_3 \over 2\pi}}
\newcommand{\dtf}{{d\th_4 \over 2\pi}}
\newcommand{\bd}{{\beta\Delta}}
\newcommand{\tn}{{1/T_1}}
\newcommand{\ot}{{O(3) NLSM }}
\newcommand{\ots}{{O(3) NLSM}}
\newcommand{\nb}{{{\bf n}}}

\section{Introduction}

The realization that one-dimensional, integer spin, antiferromagnets
possess an energy gap \cite{haldane} has made these systems the object
of intense study.
The model perhaps most commonly used to explore their properties
is the field theoretic O(3) non-linear sigma model (NLSM)
\cite{affleck,affleck1,sorensen,sagi,damle1,damle2}.
Although the model has the virtue of being integrable
\cite{zamo,zamo1}, its properties are
nonetheless only partially understood.  It is possible to access static,
thermodynamic quantities while dynamic properties, in particular,
transport properties, are in general, unavailable.  
These latter quantities depend
upon knowledge of correlation functions which are generically not exactly
computable in integrable models.  There are, of course, perturbative techniques
by which correlators in the O(3) NLSM may be analyzed.  But in strongly
coupled models, of which the O(3) NLSM is one, perturbative techniques present
a host of difficulties and so can miss qualitative (never mind quantitative)
features in the physics.

The inability to completely understand correlation functions in the
fully quantum \ot has been at the root of a recent controversy in the 
literature.  Takigawa et al. \cite{takigawa} demonstrated through measurements
of the NMR relaxation rate, $\tn$, of the Haldane gap compound, $AgVP_2S_6$,
that at long wavelengths, the spin-spin correlation functions are 
diffusive in nature.  In an elegant series of papers, Sachdev and Damle
\cite{damle1,damle2} developed a semi-classical treatment to attack the
problem and subsequently were able to describe this diffusive behaviour.
Nonetheless their computation was semi-classical leaving open the
possibility that a fully quantum treatment of the \ot would lead to
different physics.  This possibility was hinted at in the work
of Fujimoto \cite{fuji}.  There the spin conductance was computed using
exact thermodynamic considerations.  Upon subsequent work
\cite{reply}, it became clear
that the two treatments produced qualitatively different results.  In
particular, the Drude weight of the spin conductance, $D$, 
of the \ot was found to be
non-vanishing
in the zero field limit \cite{fuji}, whereas the corresponding semi-classical
treatment sees $D (H=0)=0$.  This qualitative difference opens
up the possibility that diffusive physics is not present
in the \ot itself but requires some additional
mechanism.  Such
mechanisms might include a spin-phonon coupling (as suggested by \cite{fuji}), 
spin anisotropy, inter-chain coupling (the spin-chains in $AgVP_2S_6$ are only
quasi 1-D - there do exist weak couplings in between chains
although the weakness of these couplings seems to preclude this possibility),
or perhaps small generic integrable-breaking perturbations \cite{rosch}.

In this paper we attempt to address this problem by demonstrating
a technique to compute exactly a low temperature expansion of correlators
in the O(3) NLSM.  This expansion is based upon a `form-factor' expansion.
Form-factor expansions have a long history in the computation
of correlators \cite{simo,smirbook,mus,delone,deltwo,card}.  
However these expansions have been used almost exclusively
at zero temperature.  When they have been used at finite temperature,
they have been used either in the computation of expectation
values lacking dynamical properties
\cite{leclair,salrep,delfino,mussardo} or in the development of
distinct non-perturbative representations (i.e. Fredholm determinants)
of correlators \cite{ising,leclair1} where all the terms in the expansion
were kept.
In this article we show that {\it truncated} 
form-factor expansions can be used 
to sensibly describe correlation functions at finite
temperature.  This is distinct from the programme proposed
in \cite{fred,fred1} where form factor expansions were employed
but the form factors themselves were recomputed to take direct
account of thermal fluctuations.  Here we employ the same form
factors used in zero temperature computations.

A form-factor expansion of a correlation function is predicated upon
some generic properties of integrable models.  Most importantly, the exact
eigenfunctions of the model's
fully interacting Hamiltonian are known.  With this
knowledge comes a well-defined notion of `particles' or elementary excitations
in the system.  The scattering of these particles is completely described by 
two-body S-matrices.  In particular, particle non-conserving processes are 
disallowed.  Ultimately this feature is a consequence of a series of
non-trivial conservation laws possessed by the integrable model.  In some
sense, an integrable model is a superior version of a Fermi liquid: a particle's
lifetime is infinite regardless of distance from the Fermi surface.

In order to understand these features of the O(3) NLSM, we begin by 
providing an overview of the model.  The \ot is described by the
action,
\begin{equation}\label{eIi}
S = {1\over 2g} \int dx dt (\del^\mu \nb \del_\mu \nb) ,
\end{equation}
where $\nb = (n_x,n_y,n_z)$ is a bosonic vector field constrained
to live on the unit sphere.  This action is arrived at from the Hamiltonian
of the spin chain,
\begin{equation}\label{eIii}
H = J\sum_i S_i\cdot S_{i+1} .
\end{equation}
In the continuum, large s, limit, the spin operator, $S_i$, is related
to the field, $\nb$, via
$$ 
S_i = (-1)^i s \nb_i + M_i,
$$
that is, $\nb (x,t)$ is the sub-lattice or N\'eel order
parameter.  $M$ on the other hand describes the uniform
(i.e. wavevector $k\sim 0$) magnetization.  M is related to $\nb $
via
$$
M = {1\over g} \nb \times \del_t \nb,
$$
and so is given in terms of the momentum conjugate to \nb .

The low energy excitations in the \ot take the form of a triplet of bosons.
The bosons have a relativistic dispersion relation given by
$$
E (p) = (p^2+\Delta^2)^{1/2} .
$$
Here $\Delta$ is the energy gap or mass of the bosons related to the
bare coupling, $g$, via $\Delta \sim Je^{-2\pi/g}$.  The dispersion
relation of all three bosons is identical as the model has a global
SU(2) symmetry.  The exact eigenfunctions of the \ot Hamiltonian are then
multi-particle states made up of mixtures of the three bosons.  Scattering
between the bosons is described by the S-matrix \cite{zamo}
\begin{eqnarray}\label{eIiii}
S^{a_3a_4}_{a_1a_2} (\th ) = 
\delta_{a_1a_2}\delta_{a_3a_4}\sigma_1 (\th ) &+&
\delta_{a_1a_3}\delta_{a_2a_4}\sigma_2 (\th ) +
\delta_{a_1a_4}\delta_{a_2a_3}\sigma_3 (\th );\cr\cr
\sigma_1 (\th ) &=& {2\pi i \th \over (\th + i\pi)(\th - i2\pi)};\cr
\sigma_2 (\th ) &=& {\th (\th - i\pi) \over (\th + i\pi)(\th - i2\pi)};\cr
\sigma_3 (\th ) &=& {2\pi i (i\pi - \th) \over (\th + i\pi)(\th - i2\pi)}.
\end{eqnarray}
Here $\th$ parameterizes a particle's energy/momentum via 
$E=\Delta\cosh(\th )$, $P =\Delta \sinh (\th )$.  
The primary advantage of this parameterization is
the implementation of Lorentz boosts.  Under such a boost, 
$\th \rightarrow \th +\alpha$.  As such Lorentz invariant quantities are
invariably functions of differences of rapidities.  We stress
that this relativistic invariance is a natural feature of the low
energy structure of the spin chain.  (However we do point out for
spin 1 chains, $\Delta \sim .4 J$.  As $J$ serves as the cutoff for
the theory, the low energy sector of the theory is not unambiguously
defined.)

With the excitation spectrum of the \ot in hand, we return to the form-factor
expansion.  A finite temperature expansion of correlators is given in
terms of a trace over the Boltzmann density matrix:
\begin{eqnarray}\label{eIiv}
G^{\cal O} (x,t) &=& {1\over \cal Z} 
{\rm Tr}(e^{-\beta H} {\cal O}(x,t) {\cal O}(0,0))\cr\cr
&=& {\sum_{n s_n} e^{-\beta E_{s_n}}
\langle n,s_n|{\cal O}(x,t){\cal O}(0,0)|n,s_n\rangle \over 
\sum_{n s_n} e^{-\beta E_{s_n}}
\langle n,s_n|n,s_n\rangle} .
\end{eqnarray}
Here the state, $|n,s_n\rb$, denotes a set of n-particles carrying
spin quantum numbers $\{s_n\}$.
Inserting a resolution of the identity between the two field then
leads us to a double sum,
\hskip .1in
\begin{equation}\label{eIv}
G^{\cal O} (x,t) =  {\sum_{{n s_n}\atop{m s_m}} e^{-\beta E_{s_n}}
\langle n,s_n|{\cal O}(x,t)|m,s_m\rangle
\langle m, s_m |{\cal O}(0,0)|n,s_n\rangle \over 
\sum_{n s_n} e^{-\beta E_{s_n}}
\langle n,s_n|n,s_n\rangle}.
\end{equation}
We thus have reduced the evaluation of the correlator to the evaluation of a
series of matrix elements (known as `form factors').  In an integrable
model like the \ots , these matrix elements are in principle exactly computable.
However as the number of excitations involved increases,
the functional forms of the matrix elements
become increasingly unwieldy.  This, together
with the difficultly in evaluating the sums, $\sum_{(n,s_n),(m,s_m)}$,
ensure in all but a few special cases the correlators do not admit a
closed form expression.

To surmount this we adapt an idea from zero temperature
form-factor expansions.  Rather than look at the correlator in
real space and time, we examine the (more relevant) related spectral
function, $G^\CO (k,\om )$.  In computing $G^\CO (k,\om)$, only terms
in the form factor sum with a given energy, $\om $, and momentum, $k$,
contribute to the sum,
\begin{eqnarray}\label{eIvi}
G^\CO (k,\om ) &=& {1\over \cal Z}
\sum_{{n s_n}\atop{m s_m}}\delta (\om - E_{s_n}+E_{s_m})
\delta (k - p_{s_n}+p_{s_m})\cr\cr
&&  \hskip .2in \times e^{-\beta E_{s_n}}
{\langle n,s_n|{\cal O}(0,0)|m,s_m\rangle
\langle m, s_m |{\cal O}(0,0)|n,s_n\rangle},
\end{eqnarray}
as enforced by the presence of the two delta functions.
For any $\om , k$, this dramatically reduces the number of matrix elements
one must compute\footnote{Here $G^\CO$ is simply the Fourier transform
of $G^\CO (x,t)$, but similar considerations also apply to the corresponding
retarded correlator.}.  
This reduction nonetheless leaves a difficult
computation.  However we can exploit the gapped nature of the spin chain
to make the problem more tractable.  Because the theory is gapped
(with gap, $\Delta$), the correlator admits a low temperature expansion
of the form,
\begin{equation}\label{eIvii}
G^\CO (k,\om ) = {\sum_n} \alpha_n (k,\om ) e^{-\bd} .
\end{equation}
For the particular correlators of concern in this paper and for
the range of $\om $ and $k$ in which we are interested, each $\alpha_n$
is determined by a single matrix element.  Because we can compute these
matrix elements, we obtain an {\it exact} low temperature expansion.

Our ability to compute such an expansion bears upon another controversy
in the literature.  LeClair and Mussardo \cite{leclair} argued that
it was possible to use the same form-factors we employ here to compute
finite temperature correlators.  However rather than directly evaluate
individual terms in the sum (\ref{eIv}), they first conjectured an ansatz
involving a resummation of terms in the sum.  This is described in more detail
in Section 3.  This procedure was criticized in \cite{salrep}.  There it
was argued that while this worked for the computation of one-point functions,
it was problematic for two-point functions.  Rather it was argued it
was better in general to attack such problems through the use of form factors
computed against a thermalized vacuum \cite{fred,fred1,kor}.
However the counterexample cited in \cite{salrep}, 
a computation involving interacting
quantum Hall edge states, involved a gapless theory, and so is
in a different class than the model considered in this paper.  (Without
a gap, the low temperature expansion we consider above ceases to
make sense.)  This work here shows that it is possible, at least
in certain cases, to make sense of the form-factor expansion of two
point functions at finite temperature.  
But while we can make sense of this expansion, we cannot
compare our computations directly to the ansatz posited in \cite{leclair}.
Their ansatz as is applies only to diagonal theories where scattering
does not permute internal quantum numbers, contrary to the case here.

The outline of the paper is as follows.  In Section 2 we summarize
the results of the form factor computations for three quantities:
the magnetic susceptibility, and two transport properties,
the spin conductance and the NMR relaxation
rate, $\tn$.  
The details of these computations are found in later sections or in 
appendices if the reader is so interested.  The first quantity, the
susceptibility, is compared to the susceptibility as derived from a low
temperature expansion of the exact free energy.  We see that they
match verifying our claim that the form factor expansion can yield
an exact low temperature expansion.

We compare our transport calculations to the semi-classical computations in
\cite{damle1,damle2}.  The essence of this method lies in
treating the spin-chain as a Maxwell-Boltzmann gas of spins which
interact with one another through the low energy limit of the
scattering of the \ots,
\begin{equation}\label{eIviii}
S^{cd}_{ab} (\th =0 ) = - \delta_{ad}\delta_{cb} .
\end{equation}
While static properties computed in the two treatments agree
(for the susceptibility, we find that up to temperatures on
the order of the gap, $T \sim \Delta$, the
two computations agree), we see differences in transport properties.
For the spin conductance we find, in contradistinction to the semi-classical
computation, that the Drude weight of the spin conductance
is finite in the limit 
of zero external field.  Our results for
the NMR relaxation rate, $\tn$, indicate a similar
discrepancy.  We, like \cite{sagi}, find that $\tn$ is characterized 
by ballistic logarithms.  These logarithms are relatively
robust: they continue to appear at higher orders in the low
temperature expansion.
We do not, however, see diffusive behaviour
in the relaxation rate, i.e. $\tn \sim 1/\sqrt{H}$, 
nor does our low temperature
expansion match the low temperature expansion of the semi-classical
computation of the correlator.

We consider two possibilities in explaining these discrepancies.
We argue that the structure of the conserved quantities or charges differs
between the O(3) NLSM and its semi-classical variant and that these
differences lead to ballistic behaviour on the one hand and diffusive
behaviour on the other. 
The other explanation we forward to explain this discrepancy lies in the
supra-universality of the low energy S-matrix (\ref{eIviii}).  The low
energy limit of this S-matrix is shared by generic integer spin
chains.  Indeed it is shown in \cite{damle2} that a two-leg spin-1/2
ladder, expected to share the low energy behaviour of a spin-1 chain,
has this exact low-energy S-matrix.  However rather than the supra-universality
being a virtue, it may be that it under-specifies the physics.  In this
way the semi-classical treatment, valid in and of itself (particularly
in light of its ability to reproduce experimental data), may capture
different physics than that of the \ots .  In Section 2 we consider this
further in the light of the sine-Gordon model where a similar
phenomena may be argued to occur.

In the first part of Section 3 we explain in some detail how the form-factor
expansion is to be understood.  In particular we consider the various
technical details of the expansion, including how to regulate the infinities
that appear generically in the form factors of the double expansion.
In second part of Section 3 we review the specific
form factors of the \ots .  And 
finally in Section
4, we review the low temperature expansion of the exact free energy, necessary
for comparison with the form-factor computation of the susceptibility.

\vfill\eject

\section{Summary and Discussion of Results}

\vskip .5in

\subsection{Zero Field, Finite Temperature Susceptibility}

In this subsection we present results for 
the magnetic susceptibility arising from several methods of computation:
a form factor evaluation of the magnetization-magnetization
correlator in the context of a Kubo formula, an exact computation
of the system's free energy, and finally, treating the
excitations of the $O(3)$ sigma model as non-interacting
particles obeying both a Fermi-Dirac distribution and
a Maxwell-Boltzmann distribution in the spirit of
the  semi-classical approximation
of Sachdev and Damle \cite{damle1,damle2}.  We thus will be able to 
determine the temperature regime over which
our truncation of the form factor expansion applies as well
as comparing with other computational techniques.

\subsubsection{Kubo Formula and Form Factors}

The susceptibility, $\chi$, at $H=0$ can be computed from
the magnetization-magnetization operator using a Kubo formula:
\begin{eqnarray}\label{eIIi}
\chi_(H=0) &=& C(\om = 0,k = 0) \cr\cr
C(\om = 0,k = 0) &=& \bigg[ \int^\infty_{-\infty} dx \int^\beta_0
d\tau e^{iw_n\tau}e^{ikx} 
\langle T(M^3_0(x,\tau )M^3_0(0,0))
\rangle\bigg]_{\om_n \rightarrow -i\om+\delta} .
\end{eqnarray}
To evaluate this correlator we employ an expansion in terms of
the exact eigenfunctions of the theory, i.e. a form factor expansion.
In particular we write
\begin{eqnarray}\label{eIIii}
\langle M^3_0(x,\tau )M^3_0(0,0)\rangle &=& {1\over \cal Z} 
{\rm Tr}(e^{-\beta H} {\cal O}(x,t) {\cal O}(0,0))\cr\cr
&=& {\sum_{n s_n} e^{-\beta E_{s_n}}
\langle n,S_n|{\cal O}(x,t){\cal O}(0,0)|n,S_n\rangle \over 
\sum_{n S_n} e^{-\beta E_{s_n}} \langle n,S_n|n,S_n\rangle}.
\end{eqnarray}
Here $|n,S_n\rb$ is a state of n excitations with spins described
by $S_n=\{s_1,\cdots,s_n\}$.  In writing the above we 
have suppressed sums over the energy
and momenta of the excitations.
A term in the thermal trace with n excitations is weighted by
a factor of $e^{-n\beta\Delta}$.
At low temperatures it is
thus a good approximation to truncate this trace.
For this computation we keep only terms with one and two excitations,
i.e. $n=1,2$.
To evaluate the matrix elements appearing in (\ref{eIIii}) we insert
a resolution of the identity in between the two fields.
As we only consider matrix elements involving one and two excitations
from the thermal trace, we thus have
\begin{eqnarray}\label{eIIiii}
\lb s_1 | M^3_0(x,\tau )M^3_0(0,0) | s_1 \rb 
&=& \sum_{mS_m}
\lb s_1 | M^3_0(x,\tau )|mS_m\rb
\lb mS_m | M^3_0(0,0) | s_1\rb\cr\cr
&=& \sum_{s'_1}
\lb s_1 | M^3_0(x,\tau )|s'_1\rb
\lb s'_1|M^3_0(0,0)| s_1\rb + \cdots;\cr\cr
\lb s_1 s_2|M^3_0(x,\tau )M^3_0(0,0) | 
s_2 s_1  \rb &=& \sum_{mS_m}
\lb s_1 s_2 |M^3_0(x,\tau )| m S_m\rb
\lb mS_m | M^3_0(0,0) | s_2 s_1\rb\cr\cr
&=&
\sum_{s'_1s'_2}
\lb s_1 s_2 |M^3_0(x,\tau )|s'_1s'_2\rb
\lb s'_2s'_1 | M^3_0(0,0) | s_2s_1\rb + \cdots .
\end{eqnarray}
In the above we have truncated the sum arising from the resolution
of the identity.  With the first matrix element 
of the thermal trace, we only keep
terms from the resolution of identity with one excitation.  We 
are interested in the behaviour of the susceptibility
at $\omega = 0$ and this term provides the only contribution.
Similarly, the only term arising from the second matrix element of the thermal
trace contributing to the DC susceptibility
comes from keeping the term from the resolution of the identity involving
two excitations.
Further details surrounding the methodology of this expansion and
the explicit exact evaluation of the matrix elements are found in Section
3 and Appendix A.
With such details we can evaluate $C(\om=0,k=0)$ with the
result
\begin{equation}\label{eIIiv}
C(\om = 0, k = 0) = C_1(\om = 0,k=0) + C_2(\om = 0,k=0),
\end{equation}
where $C_1$ and $C_2$ are given by
\begin{eqnarray}\label{eIIv}
C_1(\om = 0, k = 0) &=& {2\beta \Delta \over \pi} K_1(\beta\Delta);\cr\cr
C_2(\om = 0, k = 0) &=& 
-{6\beta\Delta \over \pi} K_1(2\beta\Delta ) \cr\cr
&& \hskip .1in + {2\beta \Delta \over \pi}\int d\th_1d\th_2 
e^{-\beta\Delta (\cosh(\th_1)+\cosh(\th_2))} \cosh(\th_1)
{11\pi^2+2\th_{12}^2 \over \th_{12}^4+5\pi^2\th_{12}^2+4\pi^4};\cr\cr
&=& -{6\beta\Delta \over \pi} K_1(2\beta\Delta )
+ {22 \beta \Delta \over \pi^3}K_0(\beta\Delta)K_1(\beta\Delta) 
+ {\cal O}({T\over\Delta}e^{-2\beta\Delta}),
\end{eqnarray}
where $\th_{12} = \th_1-\th_2$ and $K_n$ are standard 
modified Bessel functions.  
The first term in $C_2$ is a `disconnected' contribution
related to $C_1$.  The second term is a connected contribution and
as such is genuinely distinct from $C_1$.
We now consider such disconnected contributions
further.

\subsubsection{Resummed Form Factors}

In computing the susceptibility, we are able to go beyond the
approximation introduced in truncating the form factor sum arising
from the thermal trace.  It is possible to include `disconnected'
terms arising from higher particle contributions.  Such
disconnected terms appear when higher particle matrix elements
are evaluated.  For example when we evaluate the four excitation
matrix element $\lb s'_2s'_1 | M^3_0(0,0) | s_2s_1\rb$, we obtain
a term of the form
$$
\lb s'_2s'_1 | M^3_0(0,0) | s_2s_1\rb = \cdots + \delta_{s'_2s_1}
\lb s'_1 | M^3_0(0,0) | s_2\rb + \cdots .
$$
This term is `disconnected' in that it is directly related to a
matrix element involving a lesser number (two) of excitations.
It arises from the annihilation of $s'_2$ with $s_1$.
Such a term is responsible, as we just indicated, for the first
term of $C_2$ above.

What is remarkable is that we are able to sum up over {\it all}
possible disconnected pieces arising from arbitrarily high particle
form factors which are proportional to the connected lower particle
matrix elements already computed.  This resummation amounts to the
evaluation of a geometric series.  For example, including all disconnected
terms involving the matrix element going into the evaluation of $C_1$
modifies it as follows
\begin{eqnarray}\label{eIIvi}
C_1 = {\bd\over\pi} \int d\th e^{-\bd\cosh(\th )} \cosh (\th)
&\rightarrow& 
{\bd\over\pi} 
\int d\th e^{-\bd\cosh(\th )} \cosh (\th)
\sum_{n=0} 3^n e^{-n\bd\cosh(\th)}\cr\cr
&=& {\bd\over\pi} 
\int d\th {e^{-\bd\cosh(\th )} \cosh (\th)\over 1 + 3e^{-\bd\cosh(\th)}}\cr\cr
&=& {2\bd\over\pi} K_1(\bd ) - {6\bd\over\pi} K_1(2\bd)
+ {\cal O}(e^{-3\bd}).
\end{eqnarray}
We see resumming the disconnected pieces thus reproduces both $C_1$
and the first term in $C_2$ plus additional terms higher order
in $e^{-\bd}$.  The appearance of the factors $e^{-\bd\cosh(\th)}$
in the geometric series is natural and arises from the Boltzmann weighting
of the higher particle terms.  The combinatorial factor of 3 reflects
the three bosons in the system.

In collecting all the disconnected pieces related to the {\it connected}
term in $C_2$, we find something similar
\begin{eqnarray}\label{eIIvii}
{\rm connected}~C_2 + {\rm disconnected~terms} && \cr\cr
&& \hskip -2in = {\bd \over \pi}\int d\th_1d\th_2 
{e^{-\bd (\cosh(\th_1)+\cosh(\th_2))} \over 1+3e^{-\bd\cosh(\th_1)}}
(\cosh(\th_1)+\cosh(\th_2))
{11\pi^2+2\th_{12}^2 \over \th_{12}^4+5\pi^2\th_{12}^2+4\pi^4}.
\end{eqnarray}
One expects in general that the inclusion 
of disconnected terms from arbitrarily high
particle number will improve the accuracy of the calculation.
In the case of the susceptibility, the agreement between
the form factor computation and the exact numerical analysis
actually becomes slightly worse.  However this should not be
taken as indicative of the resummation in general.  We will
comment on this further at the end of this section.

\subsubsection{Gases of Free Particles}

For the purposes of comparison, we compute the susceptibility
of both a free electron gas (or equivalently, a system of hard-core
bosons) as well as a Maxwell-Boltzmann gas.  At sufficiently
low temperatures both
of these quantities should be close to the exact value of $\chi$ for
the $O(3)$ sigma model.  How the susceptibility of the free electron
gas deviates from the exact value of $\chi$ gives us an understanding
of the temperature at which interactions become important.  And how
the susceptibility of the Maxwellian gas deviates from the exact answer
marks the temperature at which the semi-classical approximation found
in Damle and Sachdev \cite{damle1,damle2} must begin to breakdown.

These two susceptibilities are given by
\begin{eqnarray}\label{eIIviii}
\chi_{\rm free~el.} &=& {\bd\over\pi} \int d\th 
{\cosh(\th ) e^{-\bd\cosh(\th)}\over (1+e^{-\bd\cosh(\th)})^2} \cr
&=& {2\bd\over\pi} K_1(\bd) + {\cal O}(e^{-\bd})\cr
&=& \sqrt{2\bd\over\pi}e^{-\bd}+{\cal O}({T\over\Delta}e^{-\bd});\cr\cr
\chi_{\rm MB} &=& \sqrt{2\bd\over\pi}e^{-\bd}.
\end{eqnarray}
We see that at low temperatures ($\bd \ll 1$) both of these
expressions coincide with the low temperature limit of the
form factor computation of $\chi$.  In particular,
the terms of $\CO (e^{-\bd})$ are identical.

\subsubsection{Thermodynamic Bethe Ansatz}

It is possible in the case of the $O(3)$ sigma model to arrive
at exact expressions (in the form of coupled integral equations)
for the zero-field susceptibility \cite{tsvelick,wiegmann}.  These
equations, in their most compact form, appear as
\begin{eqnarray}\label{eIIix}
\chi (H=0) &=& - {\Delta \over 2\pi} \int d\th \cosh (\th ) 
{\del^2_H \ep (\th )|_{H=0} \over 1+e^{\beta\ep(\th )}};\cr\cr
\ep (\th ) &=& \Delta\cosh(\th ) - T \int d\th'
\log (1+e^{\beta\ep_2(\th')})s(\th-\th');\cr\cr
\ep_n (\th ) &=& T \int d\th '
s(\th-\th ')\bigg\{
\log (1+e^{\beta\ep_{n-1}(\th')})+\log (1+e^{\beta\ep_{n+1}(\th ')})
+\delta_{2n}\log (1+e^{\beta\ep(\th')})\bigg\}
\cr\cr
H &=& \lim_{n\rightarrow \infty} {\ep_n(\la )\over n} 
\end{eqnarray}
We will show results from the exact numerical evaluation of these
equations in the next section.  However these equations admit
a closed form low temperature expansion.  The details of this
expansion may be found in Section 4.  Here we just give the final results
\begin{eqnarray}\label{eIIx}
\chi &=& {2\beta \Delta \over \pi} K_1(\beta\Delta)
-{6\beta\Delta \over \pi} K_1(2\beta\Delta ) \cr\cr
&& \hskip .1in + {2\beta \Delta \over \pi}\int d\th_1d\th_2 
e^{-\beta\Delta (\cosh(\th_1)+\cosh(\th_2))} \cosh(\th_1)
{11\pi^2+2\th_{12}^2 \over \th_{12}^4+5\pi^2\th_{12}^2+4\pi^4}.
\end{eqnarray}
Remarkably, we see this expansion agrees exactly with the corresponding
expression derived with the aid of form factors.  Thus
the form factor expansion at finite temperature meets an
important test.

\subsubsection{Comparison of Methodologies}

In this section we compare the various methods of computing
the susceptibility of the $O(3)$ sigma model.  In Figure 1
are plotted the susceptibilities computed via an exact numerical
analysis of the TBA equations, a low temperature expansion
of the same equations, and a computation based upon the two and
four particle form factors.
We see that as indicated previously that
the form factor computation and the low temperature expansion
match exactly.  Moreover these two computations track
the exact susceptibility over a considerable range of temperatures
despite the fact
these computations are truncated low temperature expansions.

In Figure 2 we compare both the exact TBA susceptibility and
the form factor computation of $\chi$ with the susceptibility
of a classical Maxwellian gas.
We see the results track one another
for temperatures, $T \leq \Delta$.  
For temperatures beyond $\Delta$, however,
the Maxwellian susceptibility differs markedly.
This is then roughly the temperature at which
the semi-classical approximation found in \cite{damle1,damle2}
should be expected to break down. 

In Figure 3 are plotted the exact results for the susceptibility
together with the susceptibility from the resummed form factors.  
We see that the resummed
susceptibility is somewhat higher that the exact numerics at
$T\sim 5\Delta$ and disagrees
at roughly the $10\%$ level whereas the susceptibility
computed using the unresummed form factors sees better agreement
at these same temperatures.  At lower temperatures ($T \sim 2-3\Delta$)
the disagreement between the exact numerics and the two form factor
computations is roughly the same.  In general then, the resummation
does not improve the accuracy of the computation of the susceptibility.

\vskip .5in

\subsection{Spin Conductance}

In this section we compute the spin conductivity, $\sigma_s$.
The spin conductivity gives the response of the spin chain to
a spatially varying magnetic field.  It is defined via
\begin{equation}\label{eIIxi}
j_1 (x,t) = \sigma_s \nabla H ,
\end{equation}
and so can be expressed in terms of a Kubo formula,
\begin{equation}\label{eIIxii}
{\rm Re}\thinspace\sigma_s (k,\om )= -{1\over k}
\int dx 
dte^{ikx+i\om t}\thinspace{\rm Im}\lb j_0(x,t)j_1(0,0)\rb_{\rm retarded}.
\end{equation}
In the notation used in this paper the spin current $j_1$ is synonymous
with $M_1$, the Lorentz current counterpart of the uniform magnetization,
$M_0\equiv j_0$.  We will focus primarily on computing the Drude weight, D,
of ${\rm Re}\thinspace\sigma_s$, i.e. computing the term in $\sigma_s (k,\om )$
of the form
\begin{equation}\label{eIIxiii}
\sigma_s (k=0,\om ) = D \delta (\om ).
\end{equation}
However we are able to compute $\sigma_s$ for general $k,\om$.  We
find that for $\omega \ll 2\Delta$, $k=0$, the spin conductivity is
described solely by the Drude weight.  In particular, we find no indication
of a regular contribution to $\sigma_s (k=0,\om )$.

To evaluate $\sigma_s$, we employ the identical form factor
expansion to that used in computing the susceptibility.  And like the
susceptibility, our result is an exact low temperature expansion
of $D$,
$$
D = \sum_n D_n e^{-n\bd}.
$$
Here we will compute $D_1$ and $D_2$ exactly.  As the details of the
computation are nearly identical to that of the susceptibility, we merely
write down the results:
\begin{eqnarray}\label{eIIxiv}
D(H=0) &=& {\beta \Delta} \int d\th e^{-\bd\cosh(\th)}
{\sinh^2(\th)\over\cosh(\th)}(1-3e^{-\bd\cosh(\th)}) \cr\cr
&&\hskip -.6in + 2\bd \int d\th_1d\th_2
e^{-\beta\Delta (\cosh(\th_1)+\cosh(\th_2))} 
{\sinh^2(\th_2)\over\cosh(\th_2)}
{11\pi^2+2\th_{12}^2 \over \th_{12}^4+5\pi^2\th_{12}^2+4\pi^4}
+ \CO (e^{-3\bd}) ;\cr\cr
&=& e^{-\beta\Delta} \sqrt{2\pi\over \beta\Delta} 
(1 +{\cal O}({T\over\Delta}))\cr\cr
&& \hskip .4in - e^{-2\beta\Delta} \sqrt{1\over \beta\Delta} 
({3\over 2}\sqrt{\pi} - {11\over \pi}\sqrt{T\over\Delta} 
+{\cal O}({T\over\Delta}))
+ {\cal O}(e^{-3\beta\Delta}).
\end{eqnarray}
This expression involves only the two and four particle form
factors.  If we also include all higher order disconnected terms related
to those above we find instead (akin to the susceptibility),
\begin{eqnarray}\label{eIIxv}
D(H=0) &=& {\beta \Delta} \int d\th e^{-\bd\cosh(\th)}
{\sinh^2(\th)\over\cosh(\th)}{1\over 1+3e^{\bd\cosh(\th)}} \cr\cr
&& + 2\bd \int d\th_1d\th_2
e^{-\beta\Delta (\cosh(\th_1)+\cosh(\th_2))} 
{\sinh^2(\th_2)\over\cosh(\th_2)}
{11\pi^2+2\th_{12}^2 \over \th_{12}^4+5\pi^2\th_{12}^2+4\pi^4}\nonumber\\
&& \hskip 1in \times {1\over 2}
({1\over 1+3e^{\bd\cosh(\th_1)}}+{1\over 1+3e^{\bd\cosh(\th_2)}}).
\end{eqnarray}
We plot these two results in Figure 4 as a function of $T/\Delta$.
Akin to the susceptibility, the result does not differ greatly
if the resummed disconnected terms are included.

We observe that $D(H=0) \neq 0$.  This
is in accordance with \cite{fuji} where $D$ is computed using an argument
involving the finite size scaling of 
the thermodynamic Bethe ansatz equations.  (We do note that the 
computation of $D$ at $H=0$ in \cite{fuji} appears only as a 
note added in proof and so is decidedly sketchy.  
However the equations governing $D$ developed
in \cite{fuji} are manifestly positive with the consequence $D$ cannot
vanish.)
But our results do differ from
the semi-classical computation of \cite{reply} where is was found that
$D$ vanishes at $H=0$.
And again we find no additional regular contributions
to $\sigma_s (k=0,\om =0)$ -- only the Drude term is present. 
This is true not just to the order of the form factor expansion
we work but at least to one higher order.  Moreover we are 
willing to conjecture that is true to all orders.

We have only given the spin conductivity at $H=0$.  However it is extremely
straightforward to generalize the form factor computation to finite H.
As H couples to the total spin, a conserved quantity, the form
factors, $f^\CO (x,t)$, 
of an operator, $\CO (x,t)$, carrying spin s, are altered
via the rule
$$
f^\CO (t) \rightarrow e^{iHst} f^\CO (t).
$$
(In the case of the spin conductance, the spin currents, $j_\mu=M^3_\mu$, carry
no spin and so are not altered at all.)  The only remaining change
induced by a finite field is to the Boltzmann factor appearing in the
thermal trace.  If an excitation with rapidity, $\th$, carries spin s,
its Boltzmann factor becomes
$$
e^{-\beta(\Delta\cosh (\th )-sH)}.
$$
For example we find $D$ as a function of $H$ (to $\CO(e^{-\bd})$)
to be
\begin{equation}\label{eIIxvi}
D(H) = {\beta \Delta} \cosh (\beta H) \int d\th e^{-\bd\cosh(\th)}
{\sinh^2(\th)\over\cosh(\th)} .
\end{equation}
Again this in agreement with \cite{fuji}.  Indeed \cite{fuji}
computes $D(H)$ at large $H/T$ (but $H\ll \Delta$) to be
\begin{equation}\label{eIIxvia}
D = {\beta\Delta\over 4\pi} e^{\beta H} \int d\th {\sinh^2 (\th ) \over \cosh (\th )} e^{-\beta\Delta\cosh (\th)} + {\cal O}(e^{-2\beta\Delta}).
\end{equation}
Up to a factor of $2\pi$, this expression is in exact agreement with
\ref{eIIxvi}.
In this particular case our derivation of $D(H)$ agrees
with the semi-classical computation
\cite{reply} (provided $T \ll H \ll \Delta$).  The symmetries
in the semi-classical model that lead $D(H=0)$ to vanish are broken
for finite H.

\vskip .5in

\newcommand{\obd}{{\CO (e^{-\bd})}}
\newcommand{\otbd}{{\CO (e^{-2\bd})}}

\subsection{NMR Correlators}

In this section we compute the NMR relaxation rate, $\tn$.
We are interested in computing this rate in order to compare it
to the experimental data found in \cite{takigawa} on the relaxation
rate of the quasi one-dimensional spin chain, $AgVP_2S_6$.
For temperatures in excess of $100K$ (the gap, $\Delta$,
in this compound is on the order of $320K$),
the experimental data \cite{takigawa}
shows the relaxation rate to have an inverse dependence
upon $\sqrt{H}$:
$$
\tn \propto {1\over\sqrt{H}}.
$$
This dependence is nicely reproduced by the semi-classical methodology
in \cite{damle1,damle2}.  Moreover the semi-classical computation
reproduces the activated behaviour of $\tn$ in this same
temperature regime:
$$ 
\tn \propto e^{-3\bd/2}.
$$
We are interested in determining whether a calculation in the fully
quantum O(3) NLSM can reproduce these results.  To this end we
compute $\tn$ using a form factor expansion.  Sagi and Affleck \cite{sagi}
have already done such a computation to lowest order in $e^{-\bd}$.
But they do not find the above behaviour.  Rather they see
$$
\tn \propto \log (H) ; ~~~~~ \tn \propto e^{-\bd} .
$$
We continue this computation one further step, computing to $\CO (e^{-2\bd})$.
Given the behaviour, $\tn \sim H^{-1/2}$, appears only as T is increased beyond
100K (i.e. $T/\Delta \sim 1/3$), 
it is not unreasonable to suppose higher order terms in a low
temperature expansion of $\tn$ are needed to see this singularity.

To proceed with the computation of $\tn$, we review its constituent
elements.  $\tn$ can be expressed in terms of the spin-spin
correlation function \cite{sagi}:
\begin{equation}\label{eIIxvii}
\tn = \sum_{\alpha = 1,2 \atop \beta = 1,2,3}
\int {dk\over 2\pi} A_{\alpha\beta}(k)A_{\alpha\gamma}(-k)
\lb M^\beta_0M^\gamma_0 \rb (k,\om_N),
\end{equation}
where $\om_N = \gamma_N H$ is the nuclear Lamour frequency with 
$\gamma_N$ the nuclear gyromagnetic ratio and the $A_{\alpha\beta}$
are the hyperfine coupling constants.  In the above we assume H is
aligned in the 3-direction.
The above integral is dominated
by values of $k$ near 0 \cite{sagi}.  Moreover in the relevant experiment,
the hyperfine couplings are such that only $\lb M^1_0M^1_0 \rb$
contributes.  Hence
\begin{equation}\label{eIIxviii}
\tn \propto \lb M^1_0M^1_0 \rb (x=0,\om_N \sim 0).
\end{equation}
We now proceed to compute $\lb M^1_0M^1_0 \rb$.

To compute $\lb M^1_0M^1_0 \rb$, we again employ a form factor
expansion.  Akin to the computation of the susceptibility and 
the spin conductance, this computation amounts to a low temperature
expansion of $\lb M^1_0M^1_0 \rb$,
$$
\lb M^1_0M^1_0 \rb = a_1 e^{-\bd} + a_2 e^{-2\bd} + \cdots ,
$$
where we are able to compute $a_1$ and $a_2$.
We place the details of this computations in Appendix B, here
merely quoting results:
\begin{eqnarray}\label{eIIxix}
\lb M^1_0M^1_0 \rb (x=0,\om = 0) &=& \bigg({2\Delta\over\pi}e^{-\bd}
\big(\log ({4T\over H}) - \gamma\big) -
{6\Delta\over\pi}e^{-2\bd}\big(\log ({2T\over H}) - \gamma\big)\cr\cr
&& \hskip -.8in +  {\Delta}e^{-2\bd}\big(\log ({4T\over H})-\gamma\big)
\sqrt{2\pi\over\bd}(24\pi + {17\over\pi^3})\bigg)
\big(1 + \CO (H/T) + \CO (T/\Delta)\big),
\end{eqnarray}
where $\gamma=.577\ldots$ is Euler's constant.
We are interested in the regime $H \ll T \ll \Delta$ (the
regime where it is expected spin diffusion produces singular behaviour
in $\tn$).  The terms that we have dropped do not affect this behaviour.
In principle there is no difficulty in writing down the exact expression
(to $\CO (e^{-2\bd})$); it is merely unwieldy.
This expression for $\tn$ is plotted in Figure 5 for a variety 
of values of the ratio $T/\Delta$.

We see that we do not obtain the same behaviour 
as found in \cite{damle1,damle2}.
Going to the next order in $\CO (e^{-2\bd})$ produces a behaviour in $\tn$
as $H\rightarrow 0$ identical to the lower order computation of $\obd$:
we again find a logarithmic behaviour consistent with ballistic transport.
An alternative comparison we might make to the results of \cite{damle1,damle2}
is to perform a low temperature expansion (in $\obd$) of the semi-classical
computation of $\lb M^1_0M^1_0 \rb (x=0,\om = 0)$.
Doing so by
treating $Te^{-\bd}/H$ as a small parameter, we find
\begin{equation}\label{eIIxx}
\lb M^1_0M^1_0 \rb (x=0,\om = 0) \propto \Delta e^{-\bd}
(\log ({4T\over H}) - \gamma + ({\pi\over 4} - {1\over 2}){T^2\over \pi H^2}
e^{-2\bd} + \CO (e^{-3\bd})).
\end{equation}
We see that the low temperature expansion of the
semi-classical result agrees to leading order with
our computation but afterward differs.  (We have already
seen that this occurs with the computation of the susceptibility.)
It possesses no term of $\CO(e^{-2\bd})$.  The next term rather
appears at $\CO (e^{-3\bd})$ and possesses a $1/H^2$ divergence.
That the small $H$ behaviour
should be $1/\sqrt{H}$ does suggest the importance
of summing up terms.  But the lack
of a term of $\CO (e^{-2\bd})$ in the semi-classical result nonetheless
hints that the two results are genuinely different.

\newcommand{\hb}{{\hat\beta}}

\subsection{Discussion}

We have demonstrated that it is possible to compute exact low
temperature expansions of correlators using form factors.  Moreover
we have done so in a non-trivial theory where particle scattering
sees the exchange of quantum numbers.  An important question to
answer concerns the breadth of the applicability 
of our techniques.  Our ability
to carry out these computations was partially predicated upon the 
particular correlators we studied.  For example, the fact
that only a single matrix element contributes at $\CO (e^{-\bd})$
and $\CO (e^{-2\bd})$
in the computation of the susceptibility
is related to the magnetization operator in the \ot model being a Lorentz
current density.  Because of these particular details, we thus 
expect that exact low temperature expansions of correlators
will not be available in all theories.  

The computation of correlators is done in the context of a grand
canonical partition function.  Specifically, we do not work at fixed
particle number but include matrix elements involving an arbitrary
number of particles or excitations (see \ref{eIv} for example).
This differs from the treatment found in \cite{kor}.
There correlators are computed in a canonical ensemble using
form factors at some fixed particle number, $N$.  A
thermodynamic limit is then taken, $N,L \rightarrow \infty$
holding $N/L$ (i.e. the particle density) fixed.  On a technical
level these methods may seem ostensibly different.  In particular
in this paper we end up computing correlators using form factors involving
a small finite number of particles whereas \cite{kor} computes correlators
using form factors involving a diverging number of particles.  It might
appear then that we are somehow missing information that arises in
working at a finite particle density.  
This would seem crucial in computing transport properties where 
a finite particle density is necessarily determinant.

However this difference is only apparent.  The N particle form factors
used by \cite{kor} include disconnected terms.  These disconnected terms
are equivalent to form factors involving small numbers of particles.  The
(large) N-particle form factors then contain the same information we use
in our representation of the correlators.  Moreover we can 
make this identification precise.  Our use of form factors
in the grand canonical ensemble involving some few number of particles, $n$,
is predicated upon the small parameter, $e^{-n\Delta\beta}$.  But the
disconnected terms of an N-particle form factor 
involving $n$ particles (with $n<N$) are similarly weighted
by the same small parameter, $e^{-\beta\Delta n}$.  More generally,
the presence of a gap, $\Delta$, thus means we can in principle 
create an explicit map between the two approaches.

The semi-classical method found in \cite{damle1,damle2} is similar
to the approach taken in \cite{kor} in that it uses a canonical ensemble.
It is an interesting question whether a grand canonical ensemble approach
can be developed in this same semi-classical approach.  The answer
is not obvious.  Our method works (at least at the technical level)
because we can readily identify disconnected terms.  It is not clear
whether a similar identification can be made semi-classically.

We do want to emphasize a caveat to our methodology as discussed
in some detail in Section 3.  It is unclear whether it is possible
to compute quantities that show non-analyticities as $T \rightarrow 0$.
For example it is not obvious how to compute the thermal
broadening present in the single particle spectral function.  At
$T=0$ it takes the form
\begin{equation}\label{eIIxxi}
\langle n n \rangle (\om , k) \sim \delta (\om - \sqrt{k^2+\Delta^2}) ,
\end{equation}
but is expected to broaden into a Gaussian-like peak at finite T.
To see this in a form factor expansion would likely require a resummation
of terms.  However it may well be feasible to deduce the necessary
resummation from the lower order terms in the form factor expansion.

We have also discussed using a resummation of higher order `disconnected'
terms to improve the form-factor computation.  For the quantities
considered, it turned out the resummation did not provide
a real improvement to the original computation.
Nevertheless we would guess that in general, the resummed form
factors will provide a more reliable answer as the temperature
is increased.  It is an artefact of the
above cases that they do not do so here.
For example, we 
see that at extremely high temperatures, the susceptibility
as computed by either of the form-factors methods saturates to a constant.
As such, errors in either method are cutoff -- as these expressions
at $T=\infty$ do not differ greatly from their low $T$ values, any
potential error is bounded.  If instead we computed the finite
field magnetization where we would expect a linear $T$ dependence,
the differences between the two form-factor computations would
be comparatively magnified.

To come to some sort of judgement between the form-factor and
the semi-classical approaches, an
understanding is needed of the differences between
our computations of the spin conductance and the NMR relaxation
rate.  In the case
of the first quantity, it is likely this difference is real and
not an artefact of our methodology.  The data
that goes into the spin conductance is identical to that needed to
compute the susceptibility and we know that we can match the low
temperature expansion of the susceptibility with a similar expansion
coming from the exact free energy.  Moreover we know that the Drude weight of
$\sigma_s (H=0)$ has been found to be finite from an approach
\cite{fuji} independent of ours.

In generic systems the Drude weight, $D$, of a conductivity at finite
temperatures will be zero.  It is then the integrability  of
the O(3) NLSM and the attendant existence of an infinite number of conserved
quantities that leads to a finite weight.  The existence of these
quantities can be directly related to a finite $D$.  As discussed in 
\cite{zotos}, $D$ is bounded from below via an inequality developed
by Mazur:
\begin{equation}\label{eIIxxia}
D \geq c \sum_n {\langle J Q_n \rangle \over \langle Q_n^2\rangle },
\end{equation}
where $J$ is the relevant current operator, $Q_n$ are a set of
orthogonal conserved quantities, i.e. 
$\langle Q_n Q_m \rangle = \delta_{nm}\langle Q_n^2 \rangle$, and
$c$ is some constant.
For a finite Drude weight we then require that at least one matrix
element, $\langle J Q_n \rangle$, does not vanish.  While we do no
direct computations, we can obtain an indication of whether the matrix
elements vanish by examining the symmetries of the model.
Under the discrete ($Z_2$) symmetries of the O(3) NLSM, the 
spin current, $J$, transforms via
$$
Z_2(J) \rightarrow \pm J.
$$ 
In order that the matrix element, $\langle JQ_n \rangle$, not vanish
we require that
$$
Z_2(Q_n) \rightarrow \pm Q_n.
$$ 
The $Z_2$ symmetries in the O(3) NLSM include 
$n_a \rightarrow -n_a$, $a=1,2,3$,
parity, and time reversal.  The spin current we are interested in
transforms under rotations as a vector.  Thus any charge, $Q_n$,
coupling to the current must also transform as such.  From the
work by L\"uscher \cite{lus}, it is clear there is at least one conserved
vectorial quantity such that $\langle JQ_n \rangle$ does not vanish due
to the action of one of the above $Z_2$ symmetries.

While the structure of the conserved quantities in the O(3) NLSM
seem to be consistent with the existence of a finite Drude
weight, this is not the case in the semi-classical approach.
The dynamics of the semi-classical approximation used in 
\cite{damle1,damle2} also admit an infinite number of conserved quantities
(but importantly, different than those appearing in the fully
quantum model).
However as shown there, the structure of the $Z_2$ symmetries
in the semi-classical approach is such that all matrix elements,
$\langle J Q_n \rangle$, vanish.  It would thus seem the absence
of a Drude weight in the semi-classical case is a consequence
of differences in the symmetries between the semi-classical and fully
quantum models.

To understand the discrepancies in 
the case of the NMR relaxation rate, $1/T_1$, is not as simple.
However if we believe that the spin conductance demonstrates finite 
temperature ballistic behaviour, it is hardly surprising to find
the NMR relaxation rate characterized by ballistic logarithms.  Again
the difference between the fully quantum treatment and the semi-classical
approach will lie 
in the differences between the models' conserved quantities.  Nonetheless
one possibility that we must consider is that merely going to 
$\CO (e^{-2\bd})$ in the computation of $1/T_1$
is insufficient.  It is possible that we need 
to perform some resummation of contributions from all orders to see
the desired singular behaviour, $\tn \sim 1/\sqrt{H}$.  While this would
belie our experience with computing the susceptibility and 
the spin conductance via the correlators,
the data that goes into
the two computations is not exactly identical.  Thus the possibility
that the
low temperature expansion of $\tn$ is not well controlled cannot be
entirely ruled out.

The differences in the nature of the conserved quantities between
the O(3) NLSM and the semi-classical model of \cite{damle1,damle2}
suggest the latter is not equivalent to the O(3) NLSM,
even at low energies.  An indication of this lack of equivalency
may lie in the universal nature of the ultra low energy S-matrix.
This quantity is the primary input of the semi-classical model.
The semi-classical model imagines a set of classical
spins interacting via
$$
S^{cd}_{ab} (\th = 0)  = -\delta_{ad}\delta_{cb} ,
$$
i.e. in the scattering of two spins, the spins exchange their 
quantum numbers.  However this specification may be insufficient
to adequately describe the O(3) NLSM.  Even beyond the quantum interference
effects which are neglected by the semi-classical 
treatment, it is not clear that the zero-momentum S-matrix
is enough to determine the model.  

In this light it is instructive
to consider the sine-Gordon model in its repulsive regime.  
The sine-Gordon model is given by the action,
\begin{equation}\label{eIIxxii}
S = {1\over 8\pi} \int dx dt \bigg( \partial_\mu \Phi \partial^\mu \Phi
+ \lambda \cos ({\hat\beta}\Phi )\bigg),
\end{equation}
where $\hat\beta = \beta/\sqrt{4\pi}$.
The model is generically gapped.  Its repulsive regime occurs in
the range, $4\pi < \beta^2 < 8\pi$.  The model's
spectrum then consists solely of a doublet of solitons carrying
U(1) charge.  It is repulsive in the sense that the solitons have no bound
states.
The sine-Gordon
model has a similar low energy S-matrix to the O(3) NLSM,
$$
S^{cd}_{ab} (\th = 0)  = -\delta_{ad}\delta_{cb} ,
$$
where here the particle indices range over $\pm$, the two solitons
in the theory.  Thus we might expect that sine-Gordon model
to possess identical low energy behaviour over its entire repulsive
regime. 

This is likely to be in general untrue.  
For example we might consider the behaviour
of the single particle spectral function.  We might thus want
to compute a correlator of the form
$$
\langle \psi_+ (x,t) \psi_- (0,0) \rangle ,
$$
where $\psi_\pm$ are Mandelstam fermions given by
\begin{eqnarray}\label{eIIxxiii}
\psi_\pm (x,t) &=&
\exp \bigg( \pm {i\over 2}({1\over \hb}+\hb)\phi_L (x,t)
\mp {i\over 2}({1\over \hb}-\hb)\phi_R (x,t)\bigg);\cr\cr
\phi_{L/R} &=& {1\over 2}\bigg(\Phi (x,t) 
\pm i \int^x_{-\infty} dy \partial_t \Phi (y,t)\bigg) .
\end{eqnarray}
As these fields depend explicitly upon $\hb$, it is hard
to see how the properties of the above correlator, even at
low energies could be independent of this same quantity.
More generally, $\hb$ determines the compactification radius
of the boson in the model and so is related in a fundamental
way to the model's properties.

It is useful to point out that Mandelstam fermions are the unique
fields that create/destroy solitons that carry Lorentz spin 1/2, i.e.
a spin that is independent of $\hb$.  They would then be the only
fields with a chance of matching any semi-classical computation.
However there are other soliton creation fields, for example,
$$
e^{\pm i\phi_{L,R}/\hb},
$$
for which one could determine the corresponding spectral density.
As these fields carry spin that varies as a function of $\hb$,
their spectral functions will depend upon more than the ultra low
energy soliton S-matrix.  In general, the semi-classical treatment
of the sine-Gordon model cannot capture its full quantum field content.

As with the O(3) NLSM, the conductance of the fully quantum
model differs from that of the semi-classical treatment.
If one were to compute the conductance at finite temperature in
the sine-Gordon model one would again find a finite Drude weight, $D$,
while the semi-classical approach yields $D=0$ \cite{rosch}.
The notion of under-specificity appears here again.  The semi-classical
approach for the sine-Gordon model equally 
well describes the Hubbard model at half-filling (the solitons are replaced
by particle/hole excitations in the half-filled band).  But it fails to 
give the correct Drude weight.  An analysis of finite size corrections
to the free energy in the presence of an Aharonov-Bohm flux 
\cite{fuji1} again finds a finite Drude weight in the half-filled
Hubbard model at finite temperature.

Interestingly however, there are certain properties at low energies
that seem to be independent of $\hb$.  For example, if one were to
compute the low temperature static charge susceptibility, the term
of $\CO (e^{-\bd})$ would be independent of $\hb$.  However at the
next order, $\CO (e^{-2\bd})$, this would almost certainly cease to be
true.  And the energy/temperature ranges we are interested in exploring
do not permit dropping terms of $\CO (e^{-2\bd})$.

It is important to stress we do not question the agreement between the
semi-classical model and experiment.  What we do question is whether
the fully quantum O(3) NLSM exhibits spin diffusivity.  If we are then
to understand spin diffusion in terms of the O(3) NLSM, it is possible
we need to include additional physics such as an easy axis spin
anisotropy (weakly present in the experimental system, $AgVP_2S_6$),
inter-chain couplings, or a spin-phonon coupling (as done in \cite{fuji}).

Beyond these, another mechanism that might lead to diffusive behaviour are 
small integrable breaking perturbations of the O(3) NLSM.  Generically any
physical realization of a spin chain will possess such perturbations,
even if arbitrarily small.  Such perturbations may introduce the necessary
ergodicity into the system, ergodicity that is absent in the integrable
model because of the presence of non-trivial conserved charges, and so
lead to diffusive behaviour.  As discussed in the semi-classical context
by Garst and Rosch \cite{rosch}, such perturbations introduce an additional
time scale, $T$, governing the decay of conserved quantities
in the problem.  For times, $t < T$, the behaviour of
the system is ballistic and the original conserved quantities do not decay.
For times, $t > T$, the behaviour is then diffusive.  Consequently the
Drude weight in the purely integrable model is transformed into a peak
in $\sigma (\omega )$ at $\omega \sim 1/T$.

Now the difference in the physics between the O(3) NLSM and its semi-classical
variant is not that of integrable breaking perturbations.  
As demonstrated in \cite{damle1,damle2},
their semi-classical model is classically integrable.  
However as discussed above the models do possess different conserved
charges.  It might then seem for certain transport quantities, 
the semi-classical model cures the lack of ergodicity present
in its quantum counterpart.

\vfill\eject

\section{Computation of Finite Temperature Correlators}

Here we present the general method by which we compute the correlators
at low but finite temperature and field: form factor expansions.
In the first part of this
section we consider the general form of these expansions
and why we expect them to be applicable at finite temperature.  In the
latter parts of this section, 
we review the exact expressions for the form factors in the
O(3) sigma model together with the necessary regulation of said form
factors at finite temperature.

\subsection{General Methodology}

To compute two-point correlation functions, we employ a
form factor expansion.  At finite temperature, such correlators
take the form
\begin{eqnarray}\label{eIIIi}
G^{\cal O} (x,t) &=& {1\over \cal Z} 
{\rm Tr}(e^{-\beta H} {\cal O}(x,t) {\cal O}(0,0))\cr\cr
&=& {\sum_{n s_n} e^{-\beta E_{s_n}}
\langle n,s_n|{\cal O}(x,t){\cal O}(0,0)|n,s_n\rangle \over 
\sum_{n s_n} e^{-\beta E_{s_n}} \langle n,s_n|n,s_n\rangle}
\end{eqnarray}
Here t can be real or imaginary
time and the sum $\sum_{ns_n}$ is over all possible
eigenstates of the Hamiltonian.  Each eigenstate is characterized
by the number of particles, $n$, in the state together with a set of
internal quantum numbers, $\{s_n\}$, in this case the value
of $S_z$ carried by each particle.
The form factor representation of the correlator is then arrived
at by inserting a resolution of the identity between the two fields:
\begin{equation}\label{eIIIii}
G^{\cal O} (x,t) =  {\sum_{{n s_n}\atop{m s_m}} e^{-\beta E_{s_n}}
\langle n,s_n|{\cal O}(x,t)|m,s_m\rangle
\langle m, s_m |{\cal O}(0,0)|n,s_n\rangle \over 
\sum_{n s_n} e^{-\beta E_{s_n}}
\langle n,s_n|n,s_n\rangle}.
\end{equation}
At zero temperature, the representation of $G^{\cal O}$ reduces
to one involving a single sum, $\sum_{m,s_m}$.

Thus the computation of $G^{\cal O}$ amounts to the evaluation
of a set of matrix elements.  These matrix elements can be computed
in principle for arbitrary $n,m$ from a knowledge of the two-body 
S-matrix together with various constraints coming from the
analytic dependence of the matrix elements upon energy-momentum.
However with increasing $n$ and $m$ the evaluation of these matrix
elements and the corresponding evaluation of the sums, $\sum_{n,s_n}$,
becomes increasingly arduous.

We are however in a better position when we consider the spectral
function corresponding to $G^{\cal O}$:
\begin{equation}\label{eIIIiii}
G^{\cal O} (x,\omega ) =  {\sum_{{n s_n}\atop{m s_m}} e^{-\beta E_{s_n}}
2\pi\delta(\omega - E_{s_m} + E_{s_n}) 
\langle n,s_n|{\cal O}(x,0)|m,s_m\rangle
\langle m, s_m |{\cal O}(0,0)|n,s_n\rangle \over 
\sum_{n s_n} e^{-\beta E_{s_n}}
\langle n,s_n|n,s_n\rangle} .
\end{equation}
We see then that only certain terms, those meeting the matching condition,
$\omega = E_{s_m} - E_{s_n}$, contribute to the spectral function.

In this paper we are concerned in particular with massive or gapped
theories.  Gapped theories are particularly amenable to this sort
of computation as they admit a notion of thresholds.
First imagine fixing ${n,s_n}$ in the sum above.  In a massive
theory the intermediate states have a finite energy.  In particular
in the O(3) sigma model, the energy of an m-particle state has
a minimum threshold of $m\Delta$.  And so states with $E_{s_m}$ exceeding
$\omega + E_{s_n}$ do not contribute to the sum.  For example if
$\omega + E_{s_n}$ is below the three particle threshold, $3\Delta$,
states with $m \geq 3$ do not make a contribution.

At zero temperature, i.e. $E_{s_n} = 0$, the notion of thresholds
leads to a situation where only a finite number of matrix elements
needs to be computed in order to obtain an {\it exact} result at
a given energy, $\omega$.  
In contrast, at finite temperature
we in general would need
to compute an infinite number of matrix elements in 
order to arrive at an exact result.  However here the massiveness of
the theory again comes to our aid.
With increasing $n$, the terms are
weighted with the Boltzmann factor, $e^{-\beta E_{s_n}} < e^{-\beta n\Delta}$.
Thus at temperatures small relative to the gap, $\Delta$, we expect 
in general only
the first terms to make a significant contribution.  We can thus
evaluate the correlator in a controlled
fashion, expanding it as the sum,
$$
G^\CO(x,\om ) = \sum_n c_n(x,\om ) e^{-n\bd} .
$$
Moreover while the evaluation of this
sum in its entirety would require the evaluation of an
infinite number of matrix elements, each individual coefficient, $c_n$,
depends only upon a finite number of matrix elements
(at least in the cases considered in this paper).
As such we are able to compute these coefficients exactly.

While the ability to do so results from each $c_n$ being determined
by a small, finite number of matrix elements, this feature will
not be found in all theories.
However 
form factor expansions in massive theories have in general found to
be strongly convergent \cite{simo,mus,delone,deltwo}.  
Specifically, matrix elements,
$\langle n, s_n|{\cal O}(0,0)|m, s_m\rangle$, where $n$ and $m$ are
large have been found to be relatively small.  Even in massless theories
where there are no explicit thresholds, convergence is good provided
the engineering dimension of the operator ${\cal O}$ matches its
anomalous dimension.  Thus even if each $c_n$ were determined
by a large (even infinite) number of matrix elements it would
be possible nonetheless to arrive at a reasonable approximation
for the coefficient.

There are, however, certain situations where we do not expect
to be able to truncate the sum, $\sum_{n,s_n} e^{-\beta E_{s_n}} (~~~)$.
In certain circumstances, a physical quantity will see a transition
as the limit of zero temperature is taken that is non-analytic in
nature.  To be concrete consider the single particle spectral
function of the staggered component of the spin field:
\begin{equation}\label{eIIIiv}
S(x,\om ) = \langle n(x,\om )n(0,0)\rangle .
\end{equation}
At zero temperature, we expect that for energies, $\om < 3\Delta$,
$S(x, \om )$ takes the form of a $\delta$-function:
\begin{equation}\label{eIIIv}
S(x,\om ) = c\delta (\om - \Delta ),~~~~ {\om < 3\Delta}, ~~~T=0.
\end{equation}
However at finite temperatures this $\delta$-function is broadened.
We then do not expect to be able to see this broadening unless
we evaluate the sum, $\sum_{n,s_n} e^{-\beta E_{s_n}}$, in
its entirety.  Indeed, computing $S(x, \om )$ through the
truncation of this sum at any finite $n$ leads to
\begin{equation}\label{eIIIvi}
S(x,\om ) = c\delta (\om - \Delta ) + \cdots  ~~~T > 0.
\end{equation}
Only through the resummation of the higher order terms is
the $\delta$-function replaced by a broadened peak.
However it may well be possible to guess at the resummation on the
basis of the first terms in the series.

Rather than consider such situations, we want to focus upon
quantities that possess a smooth $T \rightarrow 0$ transition.
As such consider the behaviour of the staggered field spectral
function, $S(x,\om )$, for energies below the gap $\om <\Delta$.
At $T=0$ we have
\begin{equation}\label{eIIIvii}
S(x,|\om | < \Delta ) = 0 ,
\end{equation}
while at $T\neq 0$
\begin{equation}\label{eIIIviii}
S(x,|\om | < \Delta ) = \CO(e^{-\beta\Delta}).
\end{equation}
Thus the $T\rightarrow 0$ limit behaves in a smooth fashion.

This method is markedly different than that developed in
\cite{fred,fred1,kor}.
In our method we employ the basis of
eigenstates, $|n,s_n\rangle$, that arises from the zero temperature
problem.  There a new basis is adopted that takes into direct
account the thermalization of the vacuum state.  Let
$|0_T\rangle$ be the state with a representation of the particle
content of the system in equilibrium at finite $T$ and let
$|(n,s_n)_T\rangle$ be states that are excitations above this
thermalized ground state.  (In contrast, $|n,s_n\rangle$ are
excitations above the empty vacuum state.)   With such
a basis, the correlators have the following form factor representation:
\begin{equation}\label{eIIIix}
\langle \CO (x,t ) \CO (0,0)\rangle 
= \sum_{n,s_n} \langle 0_T|  \CO (x,t )|(n,s_n)_T\rangle
\langle (n,s_n)_T| \CO (0,0)|0_T\rangle .
\end{equation}
This method involves considerable technical complications.
In general, it is a challenge to compute the new
vacuum state $|0_T\rangle $ as well as the excitations
above $|0_T\rangle $, never mind the form factors 
$\langle 0_T| \CO (x,t) | (n,s_n)_T\rangle $.  These difficulties
are only enhanced by the non-diagonal scattering present
in the O(3) sigma model, i.e. the two-body S-matrix is
other than $S^{a'b'}_{ab} = \delta_{aa'}\delta_{bb'}$,
where no internal quantum number are exchanged.
This method was developed in particular for theories that
are massless.  However in our case the theory is gapped.
It thus makes sense to exploit the control over the sum,
$\sum_{n,s_n} e^{-\beta E_{s_n}}$, that the low temperature regime
affords us.

In some sense our approach is similar to that of LeClair
and Mussardo \cite{leclair}.  There they begin with the form factor sum
as in (\ref{eIIIi}).  However they recast the sum of the thermal
trace through introducing a set of hole excitations complementary
to the particles.  Hole excitations appear naturally in terms of the
form factors.  A typical form factor that needs to be evaluated for
a finite T correlator looks as follows,
\begin{equation}\label{eIIIx}
\langle s_1,\ep_{s_1}| \CO (x,t) | s_2, \ep_{s_2}\rangle ,
\end{equation}
where we have explicitly labelled the energy of the state.
Using crossing symmetry, this matrix element can be rewritten
as 
\begin{equation}\label{eIIIxi}
\langle \CO (x,t) | s_2, \ep_{s_2}; \bar{s_1},-\ep_{s_1}\rangle,
\end{equation}
provided $\ep_1 = \ep_2$, $s_1=s_2$ does not hold.
Here $\bar{s_1}$ is the `charge conjugate' of $s_1$.
The excitation, $(\bar{s_1},-E_{s_1})$, can be thought of as a
new type of excitation, a hole.  Thus the double sum of a two point
correlator was recast in \cite{leclair} as
\begin{equation}\label{eIIIxii}
\langle \CO (x, t)\CO (0,0)\rangle 
= \sum_{m_p,s_p;m_h,s_h} \prod_p f(\ep_{s_p}) \prod_h f(\ep_{s_h})
\langle \CO (x, t)|m_p,s_p;m_h,s_h\rangle
\langle m_h,s_h;m_p,s_p| \CO (0,0)\rangle .
\end{equation}
Notice that the partition function, ${\cal Z}$, is absent from (\ref{eIIIxii})
while new factors, $\prod f$, have been added to the expression.  Each
$f(\ep_{s})$ is the occupation number of the excitation (in
this case assumed to be fermionic), $s$,
with energy $\ep$:
$$ 
f(\ep_s ) = {1\over 1+ e^{\ep_s/T}}.
$$
These modifications represent an ansatz put forward in \cite{leclair},
and are argued to come from the regulation of the matrix elements,
\begin{equation}\label{eIIIxiii}
\langle s_1,\ep_1| \CO (x,t) | s_2, \ep_2\rangle ,
\end{equation}
in the case $s_1 = s_2$, $\ep_1 = \ep_2$.

Although this ansatz is supported in the case of one point 
functions (i.e. expectation values of the energy or spin)
\cite{leclair,salrep,delfino},
it has come under criticism for the
computation of two-point functions in \cite{salrep}.  There 
the allied case of current-current correlators in
the quantum Hall edge problem at $T=0$ but finite voltage
was examined and it was found that their ansatz did not seem
to reproduce the correct results.

What relevance does this critique have for our approach?  We do
not and cannot use the ansatz of LeClair and Mussardo as scattering
in our theory is non-diagonal and their ansatz only makes sense in
the case of theories that are diagonal.  However might the critique
in \cite{salrep} still have bearing upon our results?  We do not think so.
The correlator considered in \cite{salrep} is 
computed in a massless theory whereas our
results depend upon the gapped nature of the O(3) sigma model
producing a series of thresholds.  
Moreover we already expect to run into difficulties 
whenever there is non-analytic
behaviour at $T=0$ near a threshold as in the behaviour of the
staggered field spectral function near $\omega \sim \Delta$.
Thus we do not expect to capture the physics of the conflation
of all the thresholds in a massless theory.

\subsection{Form Factors in the O(3) Sigma Model}

\subsubsection{Constraints Upon Form Factors}

The form factors of a field $\CO$ are defined as the matrix elements
of the field with some number of particles, $A_{a} (\th )$:
\begin{equation}\label{eIIIxiv}
f^{\CO}_{a_1\cdots a_n} (\th_1,\cdots ,\th_n ) = 
\langle \CO (0,0) A_{a_n} (\th_n ) \cdots A_{a_1} (\th_1)\rangle.
\end{equation}
The $A_{a} (\th )$ are Faddeev-Zamolodchikov operators which create
and destroy the elementary excitations of the theory.   $\th$ is the
rapidity which encodes the energy-momentum carried by the excitation,
\begin{equation}\label{eIIIxv}
p = \Delta \sinh (\th ); ~~~~ E = \Delta \cosh (\th ).
\end{equation}
The form of $f^{\CO}_{a_1\cdots a_n}$ is determined by a combination
of two-body scattering, Lorentz invariance, analyticity, and hermiticity.

The constraint from scattering is derived from the commutation relations
of Faddeev-Zamolodchikov operators:
\begin{eqnarray}\label{eIIIxvi}
A_{a_1} (\th_1 ) A_{a_2} (\th_2 ) &=& 
S^{a_3a_4}_{a_1a_2} (\th_1-\th_2) A_{a_4} (\th_4) A_{a_3} (\th_3);\cr\cr
A^\dagger_{a_1} (\th_1 ) A^\dagger_{a_2} (\th_2 ) &=& 
S^{a_3a_4}_{a_1a_2} (\th_1-\th_2) 
A^\dagger_{a_4} (\th_4) A^\dagger_{a_3} (\th_3);\cr\cr
A^\dagger_{a_1} (\th_1 ) A_{a_2} (\th_2 ) &=& 
\delta_{a_1a_2}\delta(\th_1-\th_2)+
S^{a_3a_1}_{a_2a_4} (\th_1-\th_2) A_{a_3} (\th_4) A^\dagger_{a_4} (\th_3).
\end{eqnarray}
S, the two-body S-matrix, 
gives the amplitude of the process by which particles
$\{a_1, a_2\}$ scatter into $\{a_3, a_4\}$.  It is solely a function of
$\th_1-\th_2 \equiv \th_{12}$ by Lorentz invariance.  
In our case, scattering between magnons in the O(3) model, the S-matrix
is given by
\begin{eqnarray}\label{eIIIxvii}
S^{a_3a_4}_{a_1a_2} (\th ) = 
\delta_{a_1a_2}\delta_{a_3a_4}\sigma_1 (\th ) &+&
\delta_{a_1a_3}\delta_{a_2a_4}\sigma_2 (\th ) +
\delta_{a_1a_4}\delta_{a_2a_3}\sigma_3 (\th );\cr\cr
\sigma_1 (\th ) &=& {2\pi i \th \over (\th + i\pi)(\th - i2\pi)};\cr\cr
\sigma_2 (\th ) &=& {\th (\th - i\pi) \over (\th + i\pi)(\th - i2\pi)};\cr\cr
\sigma_3 (\th ) &=& {2\pi i (i\pi - \th) \over (\th + i\pi)(\th - i2\pi)}.
\end{eqnarray}
As $\th \rightarrow 0$, the S-matrix reduces to 
$S^{a_3a_4}_{a_1a_2} = -\delta_{a_1a_4}\delta_{a_2a_3}$.  This is
the approximation 
underlying the semi-classical analysis of Damle and Sachdev 
\cite{damle1,damle2}.
For the form factor 
to be consistent with two body scattering we must then have
\begin{eqnarray}\label{eIIIxviii}
f^{\CO}_{a_1,\cdots ,a_{i+1},a_i,\cdots , a_n} 
(\th_1,\cdots,\th_{i+1},\th_i,\cdots,\th_n) &=& \cr\cr
&& \hskip -2.5in S^{{a'}_i,{a'}_{i+1}}_{a_ia_{i+1}} (\th_i -\th_{i+1})
f^{\CO}_{a_1,\cdots ,{a'}_i,{a'}_{i+1},\cdots , a_n} 
(\th_1,\cdots,\th_i,\th_{i+1},\cdots,\th_n).
\end{eqnarray}
This relation is arrived 
at by commuting the $i$-th and $i+1$-th particle.

A second constraint 
upon the form factor can be thought of as a periodicity
axiom.  In continuing 
the rapidity, $\th$, of a particle to $\th - 2\pi i$,
the particle's energy-momentum is unchanged.  However the form-factor
is not so invariant.  We instead have 
\begin{eqnarray}\label{eIIIxix}
f^{\CO}_{a_1,\cdots , a_n} 
(\th_1,\cdots ,\th_n) &=& 
f^{\CO}_{a_n,a_1,\cdots ,a_{n-1}} 
(\th_n-2\pi i,\th_1,\cdots,\th_{n-1}).
\end{eqnarray}
This constraint is derived from crossing symmetry \cite{leclair}.
It implicitly assumes that 
the field $\CO$ is local: if $\CO$ is non-local
additional braiding 
phases appear in the above relation \cite{double,smir,so8}.

Another condition related to analyticity that a form factor must
satisfy is the annihilation pole axiom.  This condition arises in 
form factors involving a particle and its anti-particle.
Under the appropriate analytical continuation, such a combination
of particles are able to annihilate one another.  As such this
condition relates form factors with $n$ particles to those
with $n-2$ particles.  In the case of the $O(3)$ sigma model
it takes the form
\begin{eqnarray}\label{eIIIxx}
i ~{\rm res}_{\th_n = \th_{n-1} +\pi i}
f(\th_1, \cdots, \th_n)_{a_1,\cdots ,a_n} 
&=& f(\th_1, \cdots, \th_{n-2})_{a'_1,\cdots ,a'_{n-2}} 
\delta_{a_na'_{n-1}}\cr
&& \hskip -2.5in \times \bigg( \delta^{a'_1}_{a_1}
\delta^{a'_2}_{a_2}\cdots \delta^{a'_{n-2}}_{a_{n-2}}\delta^{a'_{n-1}}_{a_{n-1}}
\cr
&& \hskip -2in - S^{a'_{n-1}a'_1}_{\tau_1 a_1}(\th_{n-11})
S^{\tau_1a'_2}_{\tau_2 a_2}(\th_{n-12})
\cdots S^{\tau_{n-4}a'_{n-3}}_{\tau_{n-3} a_{n-3}}(\th_{n-1n-3})
S^{\tau_{n-3}a'_{n-2}}_{a_{n-1} a_{n-2}}(\th_{n-1n-2})\bigg).
\end{eqnarray}
This relation as written assumes that we are normalizing our
particle states as $\lb \th | \th' \rb = 2\pi \delta(\th-\th ')$.

The form factor must also 
satisfy constraints coming from Lorentz invariance.
In general, the form factor 
of a field, $\CO$, carrying Lorentz spin, s, must 
transform under a Lorentz boost, $\th_i \rightarrow \th_i + \alpha$, via
\begin{eqnarray}\label{eIIIxxi}
f^{\CO}_{a_1\cdots a_n} (\th_1+\alpha,\cdots ,\th_n +\alpha) = 
e^{s\alpha} f^{\CO}_{a_1\cdots a_n} (\th_1,\cdots ,\th_n ).
\end{eqnarray}
The particular fields we will be interested in are the magnetization
density, $M_0(x,t)$, as well 
its corresponding conserved current, $M_1(x,t)$.  
Together
they form a Lorentz two-current.  (Here $0,1$ are
Lorentz indices.  Spin indices have been suppressed.)
As this current is topological
we may rewrite it in terms of a Lorentz scalar field, $m(x,t)$:
\begin{equation}\label{eIIIxxii}
M_\mu (x,t) = \epsilon_{\mu\nu} \del^\nu m(x,t)
\end{equation}
The form factors are then determined for the field $m(x,t)$ which
obeys (\ref{eIIIxxi}) with $s=0$ while 
the corresponding form factors of $M_\mu(x,t)$
are related to those of $m(x,t)$ by
\begin{equation}\label{eIIIxxiii}
f^{M_\mu}_{a_1\cdots a_n} (\th_1,\cdots ,\th_n) =
\epsilon_{\mu\nu} P^\nu (\th_i ) 
f^{m}_{a_1\cdots a_n} (\th_1,\cdots ,\th_n),
\end{equation}
where
$P^0 = \sum_i \Delta \cosh (\th_i )$ and 
$P^1 = \sum_i \Delta \sinh (\th_i )$.

These conditions do not uniquely specify the form factors.
It is easily seen that if $f(\th_1, \cdots, \th_n)_{a_1,\cdots ,a_n}$
satisfies these axioms then so does
\begin{equation}\label{eIIIxxiv}
f(\th_1, \cdots, \th_n)_{a_1,\cdots ,a_n} 
{P_n(\cosh(\th_{ij}))\over Q_n(\cosh(\th_{ij}))},
\end{equation}
where $P_n$ and $Q_n$ are symmetric polynomials in 
$\cosh (\th_{ij}), 1\leq i,j \leq n$, and are such that
\begin{equation}\label{eIIIxxv}
P_n|_{\th_n = \th_{n-1} + \pi i} = P_{n-2};~~~Q_n|_{\th_n = \th_{n-1} + \pi i} = Q_{n-2}.
\end{equation}
To deal with this ambiguity, we employ a minimalist axiom.
We choose $P_n$ and $Q_n$ such that $P_n/Q_n$ has the minimal
number of poles and zeros in the physical strip,
$\rm {Re} (\th ) = 0$, $0< {\rm Im} \th < 2\pi$.  Additional
poles are only added in accordance with the theory's bound
state structure, an unnecessary complication in our case as
the $O(3)$ sigma model has no bound states.  Using this minimalist
ansatz, one can determine $P_n/Q_n$ up to a constant.

To determine this constant we rely upon
the action of the conserved charge
$$
S_z = \int dx M^3_0 (x,0),
$$
upon the single particle states.  We expect
\begin{equation}\label{eIIIxxviia}
\lb \th, S_z=1 | S_z | \th', S_z=1 \rb = 2\pi \delta (\th -\th').
\end{equation}
Thus from the knowledge of the two particle form factor, we can
fix the overall normalization.
To check this normalization
we will compare the
form factor computations with the results of other techniques.
For example we will compute the magnetic susceptibility using both
form factors and the thermodynamic Bethe ansatz.  Through comparing the
$T\rightarrow 0$,$H\rightarrow 0$ results, we see that
the normalization has indeed been consistently computed.

As yet another check we can fix
the phase of this constant using hermiticity.
For this purpose it is sufficient to consider 2-particle form factors.
Hermiticity then gives us
\begin{eqnarray}\label{eIIIxxvi}
\langle \CO (0,0) A_{a_2} (\th_2) A_{a_1} (\th_1)\rangle^*
&=& \langle A^\dagger_{a_1} (\th_1) A^\dagger_{a_2} (\th_2)
\CO^\dagger (0,0)\rangle\cr
&=& \lb \CO^\dagger (0,0) A_{\bar{a_1}} (\th_1-i\pi) A_{\bar{a_2}} 
(\th_2-i\pi )\rangle,
\end{eqnarray}
where the last line follows from crossing and so
\begin{equation}\label{eIIIxxvii}
f^\CO _{a_1a_2} (\th_1,\th_2 )^* = f^{\CO^\dagger}_{\bar{a_2}\bar{a_1}}
(\th_2-i\pi ,\th_1-i\pi ).
\end{equation}

\subsection{Review of $O(3)$ Sigma Model Form Factors}

From (\ref{eIIIxxii}) and (\ref{eIIIxxiii}) it is sufficient to
give the form factors for the scalar operator, $m(x,t)$.
These have been computed by both Smirnov \cite{smirbook} 
and Balog and Niedermaier \cite{balog}.
However \cite{balog} presents them in a more amenable form, 
possible in this particular case
because of the simple structure of the S-matrix
of the $O(3)$ sigma model.

Using the axioms as presented in the previous section, \cite{balog}
thus finds for the two and four particle form factors
\begin{eqnarray}\label{eIIIxxix}
f^{m_a}_{a_1a_2}(\th_1,\th_2) &=& i {\Delta \pi^2 \over 4} \ep^{aa_1a_2}
\psi(\th_{12}), ~~~~~\psi (\th ) = {\tanh^2(\th/2)\over \th}
{i\pi + \th \over 2\pi i + \th};\cr\cr\cr
f^{m_a}_{a_1a_2a_3a_4}(\th_1,\th_2,\th_3,\th_4 )
&=& -{\pi^5 \Delta \over 8}\prod_{i<j} \psi (\th_{ij}) G^{m_a}_{a_1a_2a_3a_4};
\cr\cr
&=& -{\pi^5 \Delta \over 8}\prod_{i<j} \psi (\th_{ij})\times
\bigg( \delta^{a_4a_3}\ep^{aa_2a_1} g_1(\th_i) 
+\delta^{a_4a_2}\ep^{aa_3a_1} g_2(\th_i)\cr\cr
&& \hskip -1in + \delta^{a_4a_1}\ep^{aa_3a_2} g_3(\th_i)
 +\delta^{a_3a_2}\ep^{aa_4a_1} g_4(\th_i)
+\delta^{a_3a_1}\ep^{aa_4a_2} g_5(\th_i)
+\delta^{a_2a_1}\ep^{aa_4a_3} g_6(\th_i)\bigg);\cr\cr\cr
\left(\matrix{
g_1(\th_i ) \cr
g_2(\th_i ) \cr
g_3(\th_i ) \cr
g_4(\th_i ) \cr
g_5(\th_i ) \cr
g_6(\th_i )}\right)
&=& i\left(\matrix{
-i\pi (\ut^2+\vt^2 -i\pi\ut-i\pi\vt+2\pi^2) \cr
(\ut-i\pi)\vt(\vt-i\pi) \cr
(\ut-i\pi)(\ut+i2\pi)(i\pi-\vt) \cr
\ut\vt(3\pi i-\vt) \cr
\ut(\ut-i\pi)\vt \cr
2\pi i(i\pi -\ut)\vt \cr}\right) \cr\cr\cr
&& \hskip -1.5in + i(\wt-i\pi)\left(\matrix{
-4\pi^2 -i\pi(\ut+\vt) - (\ut-\vt)^2 \cr
-2\pi^2 - 3\pi i\vt + \vt^2 \cr
-4\pi^2 + i\pi(\ut-2\vt)-\ut^2 \cr
2\pi^2 + i\pi (\ut+2\vt ) -2\ut\vt \cr
-i\pi(2\ut+\vt)+2\ut\vt \cr
-2\pi^2 +i\pi(\ut-3\vt )}\right)
+ i(\wt-i\pi)^2 \left(\matrix{
0 \cr
0 \cr
0 \cr
-\ut \cr
\vt - 2\pi i \cr
\ut - \vt }\right)
\end{eqnarray}
\vskip .25in
\noindent We have checked that these form factors
do indeed satisfy the necessary axioms and found
that the results of \cite{balog} are without typographical
error.  The reader should note however that we use a different
particle normalization than \cite{balog} 
and so the results differ by an overall multiplicative constant.

The two particle form factor differs from that appearing
in Affleck and Weston's work \cite{affleck1} on the $O(3)$ sigma model.
The two particle form factor Affleck and Weston use is given by
$$ 
f^{M^3_0}_{a_1a_2}(\th_1,\th_2) \propto 
(\cosh(\th_1)-\cosh(\th_2))\ep^{3a_1a_2}
{\tanh(\th_{12}/2)\over \th_{12}}
{i\pi + \th_{12} \over 2\pi i + \th_{12}} .
$$
This differs from our form in that it has a different Lorentz
structure and lacks an extra factor of $\tanh (\th_{12}/2)$.
The different structure of the two particle form factor
is a result of the constraint
the annihilation pole axiom places on form factors of 
different particle numbers.  If one only computes the two
particle form factor, as done in \cite{kar},
this constraint can go unsatisfied.  However in terms of
the low energy behaviour (i.e. $\th_1,\th_2 \rightarrow 0$), 
the two forms for $f^{m_a}_{a_1a_2}$
are nearly identical.

\subsection{Regularization of Form Factors}

We end this section with a discussion of the regularization
of form factors that appear in the evaluation of thermal
correlators.  Form factors with all particles either to the
right or the left of the field such as
$$
f^\CO_{a_1,\cdots,a_n}(\th_1,\cdots,\th_n) = \langle \CO (0,0)
A_{a_n}(\th_n )\cdots A_{a_1}(\th_1 )\rangle
$$
do not pose any such problems.  However the form factors encountered
in the evaluation of finite temperature correlators are of the
form
$$
\langle A_{b_m} (\tth_m ) \cdots A_{b_1} (\tth_1 )
\CO (0,0) A_{a_n} (\th_n ) \cdots A_{a_1}(\th_1 )\rangle .
$$
To understand such an object we must contend with the possibility
that $\tth_i = \th_j$, $a_i=b_j$ for some $i,j$.  From the
algebra of the Fadeev-Zamolodchikov operators (\ref{eIIIxvi}),
we know the commutation relations involve $\delta$-functions,
i.e.
\begin{equation}\label{eIIIxxx}
A^\dagger_{a_i}(\tilde{\th}_i) A_{b_j}(\th_j) = 
2\pi \delta (\tilde{\th}_i-\th_j ) \delta_{a_ib_j}
+ \cdots .
\end{equation}
It is crucial to include the contributions of the 
$\delta$-functions to the correlators.  In particular they
contribute pieces which cancel off otherwise ill-defined
terms arising from the partition function.  To do so
we must understand the above form factor to equal
\begin{eqnarray}\label{eIIIxxxi}
\langle A_{b_m} && (\tth_m ) \cdots A_{b_1} (\tth_1 )
\CO (0,0) A_{a_n} (\th_n ) \cdots A_{a_1}(\th_1 )\rangle = \cr\cr
&& \sum_{{\{a_i\} = A_1 \cup A_2} \atop {\{b_i\} = B_1 \cup B_2}}
S_{A,A_1} S_{B,B_1} \langle B_1 | A_1\rangle
\langle B_2 |\CO (0,0) |A_2\rangle_{\rm connected}.
\end{eqnarray}
The sum in the above is over all possible subsets of 
$\{ a_i \}$ and $\{ b_i \}$.  
The S-matrix $S_{A,A_1}$ arises from the commutations necessary
to rewrite $A_{a_n}(\th_n ) \cdots A_{a_1} (\th_1 ) |0\rangle$
as $A_2 A_1 |0\rangle$ and similarly for $S_{B,B_1}$.
The matrix element
$\langle B_1 | A_1 \rangle$ is evaluated using the Fadeev-Zamolodchikov
algebra.
In this way (ill-defined) terms proportional
to $\delta (0)$ are produced but which cancel 
similarly ill-defined terms arising
from the evaluation of the partition function.

The `connected' form factor appearing in the above expression is
to be understood as follows.  Using crossing symmetry, the form
factor can be rewritten as
\begin{eqnarray}\label{eIIIxxxii}
\langle B_2 |\CO (0,0) |A_2\rangle_{\rm connected} &=&
\langle A_{b'_{i_k}} (\tth_{i_k} ) \cdots A_{b'_{i_1}} (\tth_{i_1} )
\CO (0,0) A_{a'_{j_q}} (\th_{j_q} )
\cdots A_{a'_{j_1}}(\th_{j_1} )\rangle_{\rm connected} \cr
&& \hskip -.85in = \langle \CO (0,0) A_{a'_{j_q}} (\th_{j_q} )
\cdots A_{a'_{j_1}}(\th_{j_1} )
A_{\bar{b}'_{i_k}} (\tth_{i_k} -i\pi ) 
\cdots A_{\bar{b}'_{i_1}} (\tth_{i_1}-i\pi )
\rangle_{\rm connected} \cr
&& \hskip -.85in =  f^\CO _{\bar{b}'_{i_1}\cdots \bar{b}'_{i_k}a'_{j_1}\cdots a'_{j_q}}
(\tth_{i_1}-i\pi,\cdots ,\tth_{i_k}-i\pi,\th_{j_1},\cdots 
,\th_{j_q})_{\rm connected},
\end{eqnarray}
where the last relation holds provided we do not have $\th_i = \tth_j$,
$a_i = b_j$ for any $i,j$.  If this does occur we see from the
annihilation pole axiom that the form factor is not well defined,
having a pole at $\th_i = \tth_j$.  In such cases the form factor
requires regulation. 

To regulate the form factor, we employ a scheme suggested by Balog
\cite{balog1}
and used by LeClair and Mussardo\cite{leclair}.  We define
\begin{eqnarray}\label{eIIIxxxiii}
&& f^\CO_{\bar{b}'_{i_1}\cdots 
\bar{b}'_{i_k}a'_{j_1}\cdots a'_{j_q}}
(\tth_{i_1}-i\pi+i\eta_1,\cdots ,
\tth_{i_k}-i\pi+i\eta_k,\th_{j_1},\cdots ,\th_{j_q})_{\rm connected}\cr
&& \hskip .35in = {\rm finite~piece~of} \lim_{\eta_i \rightarrow 0}
f^\CO _{\bar{b}'_{i_1}\cdots \bar{b}'_{i_k}a'_{j_1}\cdots a'_{j_q}}
(\tth_{i_1}-i\pi+i\eta_1,\cdots ,
\tth_{i_k}-i\pi+i\eta_k,\th_{j_1},\cdots ,\th_{j_q}).
\end{eqnarray}
In taking the finite piece of $f^\CO$, we discard terms proportional
to $\eta_i^{-p}$ as well as terms proportional to $\eta_i/\eta_j$.
In this way the connected piece is independent of the way
the various limits $\eta_i \rightarrow 0$ are taken.
Balog \cite{balog1} has already used this prescription to compute one point
functions and successfully compare them to TBA calculations.
In \cite{balog1} it was argued that the delta functions leading to 
such terms arise from the
use of infinite volume wavefunctions.  If such wavefunctions
are replaced instead with finite volume counterparts, the delta
functions are regulated.  For example, a pole in $\eta$ is changed
as follows
\begin{equation}
{1\over i\eta} = \int d\th {\delta (\th )\over \th + i\eta}
\rightarrow \int d\th {f(\th ) \over \th + i\eta} ,
\end{equation}
where $f(\th )$ is some sharply peaked function about $\th = 0$
which in the infinite volume limit evolves into a $\delta$-function.
However the principal value of this regularized integral
is zero.  
Thus discarding the pole terms is justified in this sense.
For terms that are ratios of infinitesimals, Balog also demonstrates
that such terms, once regularized, disappear in the infinite volume
limit.

\vfill\eject

\section{Thermodynamic Bethe Ansatz at Finite Temperature and Finite Field}

In this section we review the derivation of the equations describing
the exact free energy 
(and hence the susceptibility) of the $O(3)$ sigma model together
with its low temperature expansion.  The exact description of the 
thermodynamics of the
$O(3)$ sigma model takes the form of a set of quantization conditions
for the momenta, $p_\alpha$, 
of the excitations in the ground state.  With
$p_\alpha = \Delta \sinh (\th_\alpha)$, we have the following
condition \cite{wiegmann}:
\begin{eqnarray}\label{eVi}
e^{i\Delta\sinh (\th_\alpha)} &=& 
\prod^N_{\beta = 1} {\th_\alpha - \th_\beta + i\pi 
\over \th_\alpha - \th_\beta - i\pi }
\prod^M_{\gamma=1}{\th_\alpha - \lambda_\gamma - i\pi 
\over \th_\alpha - \lambda_\gamma + i\pi };\cr\cr
\prod^N_{\beta = 1} {\th_\beta - \lambda_\alpha + i\pi 
\over \th_\beta - \lambda_\alpha - i\pi }
&=& -\prod^M_{\gamma=1}{\lambda_\gamma - \la_\alpha + i\pi 
\over \lambda_\gamma - \lambda_\alpha - i\pi }.
\end{eqnarray}
Here $\th_\beta$ are the rapidities of the other excitations in the
ground state while the $\lambda$'s mark out spin excitations
above an originally polarized ground state.  (The Bethe ansatz
construction begins with a completely polarized ground state
of spin 1 excitations above which one then creates spin 
excitations -- marked out by the $\la$'s -- in order to give the
ground state the desired spin polarization.)
$N$ is the total number of excitations in the ground state while
$M$ is the number of spin excitations.  The quantum number, $S_z$, is
then given by $S_z = N-M$.  The total energy
of the ground state in a magnetic field is then equal to
$$
E = \Delta \sum^N_{\alpha = 1} \cosh (\th_\alpha ) - H(N-M).
$$
The analysis of these equations proceeds using the string hypothesis.
The solutions of the above equations take the form
\begin{eqnarray}
&& \th_\alpha {\rm~are~real};\cr\cr
&& \lambda^{n,k}_\alpha = \lambda^n_\alpha + i\pi(n+1-2k)/2, ~~~~ 
k=1,2,\ldots,n;
\end{eqnarray}
that is the $\lambda$'s are organized into `complexes' which share
a real part, $\lambda^n_\alpha$, the centre of the complex.

In computing the free energy, we are interested in the continuum
limit of the above equations.  To arrive at this limit, we
introduce densities per unit length, $\rho (\th )$ and $\sigma_n (\la )$,
of respectively the $\th_\alpha$'s, 
and the centers, $\lambda^n_\alpha$, of the complexes.
We further introduce particle and hole densities by writing
$\rho = \rho_h + \rho_p$ and $\sigma_n = \sigma_{nh} + \sigma_{np}$.
A particle density gives the probability that the
ground state contains an excitation at a given rapidity, $\th/\lambda$,
while the hole density gives the converse probability that
the excitation at the rapidity is not found in the ground state.
Equations describing these densities can be arrived at in
a standard fashion (see Section 8.3 of \cite{anderson} for an
analogous derivation in the case of the Anderson model):
\begin{eqnarray}\label{eVii}
\rho_p (\th ) + \rho_h (\th) &=& {\Delta \over 2\pi}\cosh (\th )
+ (s*\sigma_{2h})(\th );\cr\cr
\sigma_{mp} (\la ) + \sigma_{mh} (\la ) &=& \delta_{2m}s*\rho_p(\la )
+ s*(\sigma_{m+1,h}+\sigma_{m-1,h})(\la ),
\end{eqnarray}
where $s(x) = (\pi\cosh (x))^{-1}$ and $f*g$ denotes the convolution of
these two functions:
$$
f*g = \int d\la ' f(\la - \la ')g(\la ') .
$$
From these equations the
free energy per unit length, $\Omega$, can be derived 
(again see \cite{anderson} for 
details of an analogous derivation):
\begin{eqnarray}\label{eViii}
\Omega &=& - {T\Delta \over 2\pi} \int d\th \cosh (\th )
\log (1+e^{-\beta\ep (\th )});\cr\cr
\ep (\th ) &=& \Delta \cosh (\th ) - Ts*\log(1+e^{\beta\ep_2})(\th);\cr\cr
\ep_n (\la ) &=& Ts*\log(1+e^{\beta\ep_{n-1}})(1+e^{\beta\ep_{n+1}})(\la) + 
\delta_{2n}Ts*\log(1+e^{-\beta\ep})(\la);\cr\cr
\lim_{n\rightarrow \infty} {\ep_n \over n} &=& H.
\end{eqnarray}
Here we have expressed the free energy of the system in terms
of the dressed energies (or pseudo-energies), $\ep /\ep_n$, of the excitations.
These functions give the energetic cost of making an excitation at
a given rapidity taking into account the excitation's interactions
with the other particles in the ground state.
These equations are in agreement with \cite{tsvelick} where
they were first written down and correct typos
found in \cite{fuji}.

To derive the low temperature expansion of the free energy, we follow
\cite{tsvelick}.  We solve the above equations \ref{eViii} through
iteration.  We write for each pseudo-energy, $\ep_n$
\begin{equation}\label{eViv}
1+e^{\beta\ep_n(\th )} = \sum^\infty_{m=0} r_{nm}(T,\th).
\end{equation}
This expansion is such that $r_{nm}$ is of ${\cal O}(e^{-m\bd})$.
On the basis of (\ref{eViv}) we can write the free energy as a series
in $e^{-m\bd}$:
\begin{equation}\label{eVv}
\Omega = \sum^\infty_{m=1} c_m(T) e^{-m\bd}.
\end{equation}
We will compute the $m=1,2$ terms of this expansion.

The $m=0$ term of (\ref{eViv}) is arrived at by neglecting the term
involving $\log (1+e^{-\beta\ep})$ in the equation for $\ep_n$.
If this is done, these equations reduce to
\begin{eqnarray}\label{eVvi}
\ep_n (\la ) &=& {T\over 2}\log(1+e^{\beta\ep_{n-1}})(1+e^{\beta\ep_{n+1}});\cr\cr
\lim_{n\rightarrow \infty} {\ep_n \over n} &=& H.
\end{eqnarray}
They are then algebraic in nature and admit the following solution:
\begin{eqnarray}\label{eVvii}
1+e^{\beta\ep_n} &=& r_{n0} = \phi^2(n);\cr\cr
\phi(n) &=& {\sinh ({H\over 2T}(n+1))\over \sinh ({H\over 2T})}.
\end{eqnarray}
At this order of the iteration, $\ep (\th )$ becomes
\begin{equation}\label{eVviii}
\ep (\th ) = \Delta\cosh (\th ) - T\log\phi(2),
\end{equation}
and so
\begin{equation}\label{eVix}
\Omega = -{T\Delta \over 2\pi} \int d\th\cosh (\th ) 
\log(1+3e^{-\bd\cosh (\th)}).
\end{equation}
Clearly $\Omega$ is of $\CO (e^{-\bd})$.

The next coefficient in the series (\ref{eViv}), $r_{n1}$, is found
by substituting (\ref{eVviii}) into the equations for $\ep_n$:
\begin{eqnarray}\label{eVx}
\ep_n &=& T s*\log(1+e^{\beta\ep_{n-1}})(1+e^{\beta\ep_{n+1}})
+ \delta_{2n}s*\log(1+\phi(2)e^{-\bd\cosh(\th) });\cr\cr
\lim_{n\rightarrow \infty} {\ep_n \over n} &=& H.
\end{eqnarray}
To the order in $e^{-\bd}$ to which we are working, these equations reduce 
to
\begin{equation}\label{eVxi}
{\phi^2(n) \over \phi(n-1)\phi(n+1)} r_{n1}
= (r_{n-1,1}+r_{n+1,1})*s + \delta_{2n}T \log(1+e^{-\beta\Delta\cosh (\th)})
* s.
\end{equation}
As can be directly checked, they admit the solution
\begin{eqnarray}\label{eVxii}
r_{n1} &=& {\phi(1) \over \phi(2)\phi(n)}(\phi(n+1)a_n-\phi(n-1)a_{n+2})
*s^{-1}*T\log(1+\phi(2)e^{-\bd\cosh(\th )});\cr\cr
a_n (x) &=& {2n \over 4x^2 + n^2\pi^2}.
\end{eqnarray}
With this, $\ep(\th )$ to ${\cal O}(e^{-\bd})$ takes the form
\begin{equation}\label{eVxiii}
\ep (\th ) = \Delta \cosh (\th ) - T\log\phi(2)
-T {\phi(1)\over\phi(2)}(\phi(3)a_2-\phi(1)a_4)*e^{-\bd\cosh(\th )}
+ {\cal O}(e^{-2\bd}).
\end{equation}
We can continue this procedure, obtaining $r_{nm}, m\geq 2$.  Indeed
\cite{tsvelick} goes on to compute $r_{n2}$ and so corrections
of ${\cal O}(e^{-2\bd})$ to $\ep (\th )$.

The zero field susceptibility is given by
\begin{equation}\label{eVxiv}
\chi (H=0) = - \del^2_H \Omega |_{H=0} = -{\Delta \over 2\pi}
\int d\th \cosh (\th ) 
{\del^2_H \ep (\th ) \over 1 + e^{\beta\ep (\th )}}\bigg|_{H=0}.
\end{equation}
Using $\ep (\th )$ in (\ref{eVxiii}) and expanding the above expression
to ${\cal O}(e^{-2\bd})$, we find
\begin{eqnarray}
\chi (H=0) &=& {\bd \over \pi T}\int d\th \cosh (\th ) e^{-\bd\cosh (\th )}
(1-3e^{-\bd\cosh (\th )})\cr\cr
&& + {2\bd \over \pi} \int d\th_1d\th_2 \cosh (\th_1 ) 
e^{-\bd(\cosh (\th_1 )+\cosh(\th_2))}
{2\th_{12} + 11\pi^2 \over \th_{12}^4 + 5\pi^2\th_{12}^2 + 4\pi^4}
+ {\cal O}(e^{-3\bd}),
\end{eqnarray}
where $\th_{12} = \th_1-\th_2$.  This agrees exactly with the
derivation of $\chi$ coming from the computation of the two and
four particle form-factors.

\vfill\eject

\section{Acknowledgements}

RMK acknowledges the support of the NSF through
grant number DMR-9802813.  RMK would like to also acknowledge
useful discussions with Subir Sachdev, Andr\'{e} LeClair,
Guiseppe Mussardo, Hubert Saleur, and Paul Fendley.
RMK also thanks Paul Fendley
for performing the numerical analysis of the exact TBA equations.

\vfill\eject

\appendix

\newcommand{\cor}{{\langle M^3_0(x,\tau ) M^3_0 (0,0) \rangle}}
\newcommand{\mm}{{M^3_0(x,\tau ) M^3_0 (0,0)}}
\newcommand{\mo}{{M^3_0(x,\tau )}}
\newcommand{\mt}{{M^3_0(0,0)}}
\newcommand{\eb}{{e^{-\beta E_{s_n}}}}
\newcommand{\ebt}{{e^{-\beta \Delta \cosh (\th )}}}
\newcommand{\ebtot}{{e^{-\beta \Delta (\cosh (\th_1 )+\cosh(\th_2))}}}

\newcommand{\dh}{{2\pi\delta (0)}}

\section{Computation of Magnetic Susceptibility Using Form Factors}

To compute the correlator, $\cor$, we first consider the
action of the thermal trace:
\begin{equation}\label{eAi}
C(x,\tau ) = \cor = {\sum_{s_n,n} \eb \lb n,s_n | \mm | n,s_n\rb
\over \sum_{s_n,n} \eb \lb n,s_n | n,s_n\rb} .
\end{equation}
Keeping the first two terms leads us to
\begin{eqnarray}\label{eAii}
\cor &=& \bigg(\int \dt 
\ebt \sum_a \lb A_a (\th ) | \mm | A_a (\th )\rb\cr
&& \hskip -1.25in + {1\over 2}\int \dto \dtt \ebtot \times \cr
&& \hskip .25in \sum_{a_1a_2} 
\lb A_{a_1}(\th_1)A_{a_2}(\th_2) |\mm | 
A_{a_2}(\th_2)A_{a_1}(\th_1) \rb \bigg) \cr
&& / \bigg( 1 + \sum_a\int \dt \ebt \lb A_a (\th ) | A_a (\th ) \rb \bigg)
\end{eqnarray}
Expanding the denominator then gives us
\begin{eqnarray}\label{eAiii}
C(x,\tau ) &=& 
\int \dt 
\ebt \sum_a \lb A_a (\th ) | \mm | A_a (\th )\rb \times \cr
&& \hskip .5in
\bigg( 1 - \sum_a \int \dt \ebt \lb A_a (\th ) | A_a (\th ) \rb \bigg) \cr\cr
&& +  {1\over 2}\int \dto \dtt \ebtot \times \cr
&& \hskip .25in \sum_{a_1a_2} 
\lb A_{a_1}(\th_1)A_{a_2}(\th_2) |\mm | 
A_{a_2}(\th_2)A_{a_1}(\th_1) \rb .
\end{eqnarray}
The term arising from the partition function is ill-defined as
the state normalization is given by 
$\lb A_{a} (\th ) |A_{a_1}(\th_1 ) \rb = 2\pi \delta_{aa_1}
\delta (\th - \th_1)$.
However this term will be canceled by disconnected terms arising from 
$\lb A_{a_1}(\th_1)A_{a_2}(\th_2) |\mm | A_{a_2}(\th_2)A_{a_1}(\th_1) \rb$.

To evaluate this expression we begin by computing the first term of the
trace $\lb A_a (\th ) | \mm | A_a (\th )\rb$ by inserting a resolution
of the identity between the $M^3$'s:
\begin{eqnarray}\label{eAiv}
\lb A_a (\th ) | \mm | A_a (\th )\rb &=& \sum^\infty_{n=1}
\sum_{a_1,\cdots ,a_n} {1\over n!}\int \dto \cdots \int {d\th_n\over 2\pi} \cr
&& \hskip -2in \times \lb A_a(\th)|\mo|A_{a_n}(\th_n )\cdots A_{a_1} (\th_1 )\rb 
\lb A_{a_1} (\th_1 ) \cdots A_{a_n} (\th_n )|\mt|A_a (\th )\rb.
\end{eqnarray}
We only need to keep the lowest order term, $n=1$, in this expansion;
{\it all other terms make no contribution to the susceptibility}.
The terms corresponding to n even are identically zero (by parity);
the remaining $n>1$ odd terms vanish in the low energy-low momentum
limit of the corresponding spectral function.  
(We will return to this is a moment.)  Given that we can thus
compute the entire contribution $\lb A_a (\th ) | \mm | A_a (\th )\rb$
makes to the susceptibility, we will able to find an exact correspondence
between the form factor computation and a low temperature expansion
of the exact free energy.  With the $n=1$ term
we then have
\begin{eqnarray}\label{eAv}
&& \lb A_a (\th ) | \mm | A_a (\th )\rb = 
\sum_{a_1} \int \dto\lb A_a(\th)|\mo| A_{a_1} (\th_1 )\rb \cr
&& \hskip 3.5in \times \lb A_{a_1} (\th_1 )|\mt|A_a (\th )\rb ;\cr\cr
&&\hskip .25in = 
\sum_{a_1} \int \dto e^{-\tau \Delta (\cosh (\th_1 ) -\cosh (\th ))
+ ix\Delta (\sinh (\th_1 ) -\sinh (\th ))} \cr
&& \hskip 1in \times \lb \mt| A_{a_1}(\th_1) A_{a} (\th - i\pi )\rb
\lb \mt| A_a(\th ) A_{a_1} (\th_1 - i\pi )\rb ;\cr\cr
&& \hskip .25in =
 \sum_{a_1} \int \dto e^{-\tau \Delta (\cosh (\th_1 ) -\cosh (\th ))
+ ix\Delta (\sinh (\th_1 ) -\sinh (\th ))} 
f^{M^3_0}_{aa_1}(\th -i\pi,\th_1 ) f^{M^3_0}_{a_1a}(\th_1-i\pi,\th ).
\end{eqnarray}
We have used crossing symmetry in the second line.  From Section 3.3,
the form factor $f^{M^3_0}_{aa_1}(\th ,\th_1 )$ is
given by
\begin{equation}\label{eAvi}
f^{M^3_0}_{aa_1}(\th ,\th_1 ) = 
i  {\pi^2\Delta \over 4} \ep^{3aa_1}(\sinh (\th )+\sinh(\th_1))\psi(\th-\th_1).
\end{equation}
Then the lowest order contribution, $C_1(x,\tau)$, to
the spin-spin correlator, $C(x,\tau )$, is given by
\begin{eqnarray}\label{eAvii}
C_1(x,\tau ) = 
-2 \int \dt \int \dto \ebt && e^{-\tau \Delta (\cosh (\th_1 ) -\cosh (\th ))
+ ix\Delta (\sinh (\th_1 ) -\sinh (\th ))} \cr
&& \times f^{M^3_0}_{21}(\th -i\pi,\th_1 ) f^{M^3_0}_{21}(\th_1-i\pi,\th ).
\end{eqnarray}
Fourier transforming in $x$ and $\tau$ and continuing
$\om_n \rightarrow -i\om + \delta $ yields,
\begin{equation}\label{eAviii}
C_1(\omega = 0, k = 0) =  {\beta \Delta \over \pi}
\int d\th \cosh (\th ) e^{-\beta \Delta \cosh (\th )} = 
{2\beta\Delta \over \pi}K_1(\beta\Delta),
\end{equation}
where $K_1$ is a modified Bessel function.
This has the expected small temperature behaviour,
$C_1(\omega = 0, k= 0) \sim T^{-1/2} e^{-\beta \Delta} $.

Let us consider further why the above computation gives
the sole contribution to 
$\lb A_a (\th ) | M^3_0(x)M^3_0(0) | A_a (\th )\rb $.  The next
potential contribution to this matrix element takes the form
\begin{eqnarray}\label{eAviiia}
\int d\th_1 d\th_2 d\th_3 &&
\lb A_a (\th ) | M^3_0 (x)| A_{a_1}(\th_1) A_{a_2}(\th_2)
A_{a_3} (\th_3) \rb \cr\cr
&& \hskip .1in \lb A_{a_3}(\th_3) A_{a_2}(\th_2)
A_{a_1} (\th_1) | M^3_0 (0) | A_a (\th )\rb .
\end{eqnarray}
Upon evaluation this expression produces two types of terms.  The first
is associated with the disconnected pieces of the matrix
elements appearing in the above.  An example of this type of
term is 
\begin{eqnarray}\label{eAviiib}
&& \int d\th_1 d\th_2 
\lb M^3_0 (x)| A_{a_1}(\th_1) A_{a_2}(\th_2) \rb
\lb A_{a}(\th) A_{a_2}(\th_2) 
A_{a_1} (\th_1) | M^3_0 (0) | A_a (\th )\rb \cr\cr
&& \hskip .1in = \int d\th_1 d\th_2 e^{i\Delta x(\sinh (\th_1)+\sinh (\th_2))}
(\sinh (\th_1)+\sinh(\th_2))\times \big({\rm term~regular
~in~} \th_1 {\rm ~and~} \th_2\big).
\end{eqnarray}
As $M^3_0$ is a Lorentz current, the term $(\sinh (\th_1) + \sinh(\th_2))$
appears in the above.  Thus when the Fourier transform, $\int e^{ikx}$,
is taken followed by the limit, $k\rightarrow 0$, this term vanishes
identically.

The second type of term we must deal with in evaluating (\ref{eAviiia})
takes the form
\begin{eqnarray}\label{eAviiic}
&& \int d\th_1d\th_2d\th_3 
e^{ix\Delta(\sinh(\th_1)+\sinh(\th_2)+\sinh(\th_3)-\sinh(\th))}
f^{M^3_0}_{\bar{a}a_3a_2a_1}(\th-i\pi+i\ep,\th_3,\th_2,\th_1) \cr
&& \hskip 1.55in 
\times 
f^{M^3_0}_{\bar{a_1}\bar{a_2}\bar{a_3}a}(\th_1-i\pi+i\ep_1,
\th_2-i\pi+i\ep_2,\th_3-i\pi+i\ep_3,\th),
\end{eqnarray}
and arises from the connected pieces of the matrix elements appearing
in (\ref{eAviiia}).  To evaluate this term we deform\footnote{In 
doing so we assume that time is real, not imaginary.  This does
not pose a problem as we could as well
directly evaluate the retarded correlators
as opposed to evaluating them via an analytical continuation of
imaginary time-ordered correlators.}
the contours
$\th_{1,2,3}$ via
$$
\th_{1,2,3} \rightarrow \th_{1,2,3} +i\pi .
$$
In doing so we deform through a number of poles whose residues we will
pick up.  Evaluating these residues we 
again obtain something of the form (\ref{eAviiib}).
As such, Fourier transforming and taking the $k=0$ limit
forces the term to vanish and (\ref{eAviiia}) ends up making no
contribution to the susceptibility.  In the same way, it 
is easy then to convince
oneself that terms involving an even greater number of particles
similarly do not contribute to the static susceptibility.

We now go ahead and compute the second term arising from
performing the thermal trace,
$\lb A_{a_1}(\th_1)A_{a_2}(\th_2) |\mm | 
A_{a_2}(\th_2)A_{a_1}(\th_1) \rb$.
We evaluate it as before by inserting a resolution of the
identity between the two fields.  In this case the only term
that contributes is the $n=2$
term:
\begin{eqnarray}\label{eAix}
\lb A_{a_1}(\th_1)A_{a_2}(\th_2) |\mm | 
A_{a_2}(\th_2)A_{a_1}(\th_1) \rb &=& \cr\cr
&& \hskip -2.5in {1\over 2}\sum_{a_3a_4} \int \dttr \dtf 
\lb A_{a_1}(\th_1)A_{a_2}(\th_2) | \mo | A_{a_3}(\th_3 ) A_{a_4}(\th_4 )\rb 
\cr\cr
&& \hskip -2.5in \times
\lb A_{a_4}(\th_4)A_{a_3}(\th_3) | \mt | A_{a_2}(\th_2 ) A_{a_1}(\th_1 )\rb.
\end{eqnarray}
Allowing for the presence of disconnected terms, 
the matrix elements in the above expression take the form
\begin{eqnarray}\label{eAx}
\lb A_{a_1}(\th_1)A_{a_2}(\th_2) | \mo | A_{a_3}(\th_3 ) A_{a_4}(\th_4 )\rb 
&=& \cr\cr
&& \hskip -2in \delta_{a_1a_4} 2\pi\delta(\th_1-\th_4)
f^{M^3_0}_{\bar{a}_2,a_3}(\th_2-i\pi,\th_3) \cr\cr
&& \hskip -2in +\delta_{a'_3a'_2} 2\pi \delta(\th_3-\th_2)
S^{a'_1a'_2}_{a_1a_2}(\th_{12})S^{a'_3a'_4}_{a_3a_4}(\th_{34})
f^{M^3_0}_{\bar{a}'_1,a'_4}(\th_1-i\pi,\th_4)\cr\cr
&& \hskip -2in + \delta_{a'_2a_4} 2\pi \delta(\th_2-\th_4)
S^{a'_1a'_2}_{a_1a_2}(\th_{12})
f^{M^3_0}_{\bar{a}'_1,a_3}(\th_1-i\pi,\th_3)\cr\cr
&& \hskip -2in + \delta_{a_1a'_3} 2\pi \delta(\th_1-\th_3)
S^{a'_3a'_4}_{a_3a_4}(\th_{34})
f^{M^3_0}_{\bar{a}_2,a'_4}(\th_2-i\pi,\th_4)\cr\cr
&& \hskip -2in + f^{M^3_0}_{\bar{a}_2,\bar{a}_1,a_4,a_3}
(\th_2-i\pi,\th_1-i\pi,\th_4,\th_3)_{\rm c},
\end{eqnarray}
where $f_{\rm c}$ refers to a connected form-factor.
We now substitute (\ref{eAx}) into (\ref{eAix}) and 
obtain the following after some lengthy but straightforward algebra
\begin{eqnarray}\label{eAxi}
{1\over 4} \sum_{a_1a_2a_3a_4} \int \dto \dtt \dttr \dtf 
\lb A_{a_1}(\th_1)A_{a_2}(\th_2) | \mo | A_{a_3}(\th_3 ) A_{a_4}(\th_4 )\rb \cr
&& \hskip -2.5in \times
\lb A_{a_4}(\th_4)A_{a_3}(\th_3) | \mt | A_{a_2}(\th_2 ) A_{a_1}(\th_1 )\rb
\cr\cr
&& \hskip -4.5in \equiv C_{21} + C_{22} + C_{23} + C_{24} + C_{25} 
+ C_{26};\cr\cr\cr
&& \hskip -5in 
C_{21} = 2\pi \delta (0) \sum_{a_1a_2a_3} \int \dto \dtt \dttr 
e^{-\beta\Delta(\cosh(\th_1)+\cosh(\th_2))}\cr
&& \hskip -2.5in \times
e^{-\tau\Delta (\cosh(\th_3)-\cosh(\th_2)) 
-ix\Delta(\sinh(\th_3)-\sinh(\th_2))}\cr
&& \hskip -2.5in \times f^{M^3_0}_{\bar{a}_3,a_2}(\th_3-i\pi,\th_2)
f^{M^3_0}_{\bar{a}_2,a_3}(\th_2-i\pi,\th_3);\cr\cr
&& \hskip -5in C_{22} = - 3
\int \dto \dtt 
e^{-2\beta\Delta\cosh(\th_1)}
e^{-\tau\Delta (\cosh(\th_2)-\cosh(\th_1)) -ix\Delta(\sinh(\th_2)-\sinh(\th_1))}\cr
&& \hskip -2.5in \times \sum_{a_1a_2} f^{M^3_0}_{\bar{a}_1,a_2}(\th_2-i\pi,\th_1)
f^{M^3_0}_{\bar{a}_2,a_1}(\th_1-i\pi,\th_2);\cr\cr 
&& \hskip -5in C_{23} = - 3
\int \dto \dtt 
e^{-\beta\Delta(\cosh(\th_1)+\cosh(\th_2))}
e^{-\tau\Delta(\cosh(\th_1)-\cosh(\th_2)) -ix\Delta(\sinh(\th_1)-\sinh(\th_2))}\cr
&& \hskip -2.5in \times \sum_{a_1a_2} f^{M^3_0}_{\bar{a}_1,a_2}(\th_1-i\pi,\th_2)
f^{M^3_0}_{\bar{a}_2,a_1}(\th_2-i\pi,\th_1);\cr\cr 
&& \hskip -5in 
C_{24} = {1 \over 4}\sum_{a_1a_2a_3} \int \dto \dtt \dttr 
e^{-\beta\Delta(\cosh(\th_1)+\cosh(\th_2))}\cr
\hskip -2.5in \times
e^{-\tau\Delta(\cosh(\th_3)-\cosh(\th_2)) 
-ix\Delta(\sinh(\th_3)-\sinh(\th_2))}\cr\cr
&& \hskip -4.5in \times 
\Bigg\{ f^{M^3_0}_{\bar{a}_3,\bar{a}_1,a_1,a_2}
(\th_3-i\pi,\th_1-i\pi,\th_1,\th_2)_{\rm c}
f^{M^3_0}_{\bar{a}_2,a_3}(\th_2-i\pi,\th_3)\cr
&& \hskip -4.3in +
\sum_{a_4a'_1a'_2a'_4} S^{a'_4a'_2}_{a_4a_3}(\th_{21}) 
S^{a'_2a'_1}_{a_2a_1}(\th_{13})
f^{M^3_0}_{\bar{a}_2,\bar{a}_1,a_4,a_3}
(\th_1-i\pi,\th_3-i\pi,\th_2,\th_1)_{\rm c}
f^{M^3_0}_{\bar{a}'_4,a'_1}(\th_2-i\pi,\th_3)\cr
&& \hskip -2.5in +~(\th_2 \leftrightarrow \th_3)\Bigg\};\cr\cr
&& \hskip -5in 
C_{25} = {1 \over 4}\sum_{a_1a_2a_3} \int \dto \dtt \dttr 
e^{-\beta\Delta(\cosh(\th_1)+\cosh(\th_2))}\cr
\hskip -2.5in \times
e^{-\tau\Delta(\cosh(\th_3)-\cosh(\th_2)) 
-ix\Delta(\sinh(\th_3)-\sinh(\th_2))}\cr\cr
&& \hskip -4.5in \times 
\Bigg\{ 
\sum_{a_4a'_4} S^{a'_4a_1}_{a_4a_3}(\th_{21})
f^{M^3_0}_{\bar{a}_2,\bar{a}_1,a_4,a_3}
(\th_3-i\pi,\th_1-i\pi,\th_2,\th_1)_{\rm c}
f^{M^3_0}_{\bar{a}'_4,a_2}(\th_2-i\pi,\th_3)\cr
&& \hskip -4.3in +
\sum_{a_4a'_1} S^{a_4a'_1}_{a_2a_1}(\th_{13})
f^{M^3_0}_{\bar{a}_2,\bar{a}_1,a_4,a_3}
(\th_1-i\pi,\th_3-i\pi,\th_1,\th_2)_{\rm c}
f^{M^3_0}_{\bar{a}_3,a'_1}(\th_2-i\pi,\th_3)\cr
&& \hskip -2.5in +~(\th_2 \leftrightarrow \th_3)\Bigg\};\cr\cr
&& \hskip -5in 
C_{26} = {1\over 4}\sum_{a_1a_2a_3a_4} \int \dto \dtt \dttr \dtf
e^{-\beta\Delta(\cosh(\th_1)+\cosh(\th_2))}\cr\cr
&& \hskip -4.25in 
\times e^{-\tau\Delta(\cosh(\th_3)+\cosh(\th_4)-\cosh(\th_1)-\cosh(\th_2)) 
-ix\Delta(\sinh(\th_3)+\sinh(\th_4)-\sinh(\th_1)-\sinh(\th_2))}\cr
&& \hskip -4.25in \times 
f^{M^3_0}_{\bar{a}_3,\bar{a}_4,a_1,a_2}
(\th_3-i\pi,\th_4-i\pi,\th_1,\th_2)_{\rm c}\cr
&& \hskip -4.25in \times 
f^{M^3_0}_{\bar{a}_2,\bar{a}_1,a_4,a_3}
(\th_2-i\pi,\th_1-i\pi,\th_4,\th_3)_{\rm c} .
\end{eqnarray}
Although appearing exceedingly complicated, these terms
dramatically simplify once we Fourier transform.

The first term, $C_{21}$, on the r.h.s. of (\ref{eAxi}) involves
$\delta (0)$ and so is ill-defined.  However it precisely
cancels the term arising from the evaluation of the
partition function in (\ref{eAiii}),
\begin{eqnarray}\label{eAxii}
\int \dto \dtt \sum_{a_1a_2}
e^{-\beta\Delta(\cosh(\th_1)+\cosh(\th_2))} 
\lb A_{a_1} (\th ) | \mm | A_{a_1} (\th )\rb 
\lb A_{a_2} (\th ) | A_{a_2} (\th ) \rb
\end{eqnarray}
as is evident if a resolution of the identity is inserted between
the two fields, $M^3_0$, in the above and then truncated at the one-particle
level.

Having canceled off the $\delta (0)$-terms we now look
at terms that make a genuine contribution to the spin-spin
correlator.  We first consider the completely
disconnected terms.
Fourier transforming $C_{22}$ and $C_{23}$ in 
time and space and then analytically continuing,
$\omega_n \rightarrow -i\omega + \delta$,
leads to
\begin{eqnarray}\label{eAxiii}
C_{22}(\omega = 0, k = 0) + C_{23}(\omega = 0,k=0)&=& -3
{\beta \Delta \over \pi}
\int^\infty_{-\infty} d\th \cosh (\th ) 
e^{-2\beta \Delta \cosh (\th )} \cr\cr
&=& -6 {\beta\Delta\over\pi} K_1(2\beta\Delta),
\end{eqnarray}
where again $K_1$ is a standard Bessel function.

To compute $C_{24}$ we need to evaluate the connected 
four-particle form factor.  To do so we add small
imaginary pieces to the rapidities where potential poles lurk
and take only the finite piece.  For example the first
term of $C_{24}$ upon Fourier transforming reduces to
\begin{eqnarray}\label{eAxiv}
C_{24}(\omega = 0, k = 0) &=& 
-{\beta \over 8\pi^2\Delta}\int d\th_1 d\th_2 
e^{-\beta\Delta(\cosh (\th_1)+\cosh(\th_2))} \cosh^{-1} (\th_2)\cr\cr
&& \hskip -1.25in \times
f^{M^3_0}_{21}(\th_2-i\pi,\th_2) 
\sum_{a_1} f^{M^3_0}_{1\bar{a}_1a_12}
(\th_2-i\pi,\th_1-i\pi,\th_1,\th_2)_{\rm c} ~+~ {\rm three~other~terms.}
\end{eqnarray}
Then to evaluate the connected form factor in this expression we write
\begin{eqnarray}\label{eAxv}
f^{M^3_0}_{1\bar{a}_1a_12}
(\th_2-i\pi,\th_1-i\pi,\th_1,\th_2)_{\rm c}
&=& \cr
&& \hskip -1in {\rm finite~part~of}~f^{M^3_0}_{1\bar{a}_1a_12}
(\th_2-i\pi,\th_1-i\pi,\th_1-i\eta,\th_2-i\delta)
\end{eqnarray}
We evaluate this matrix element using the discussion in
Section III.D, throwing away any poles in $\eta$ or $\delta$
together with terms of the form $\eta/\delta$.
Expanding the form factor on the r.h.s. of (\ref{eAxv}) in $\eta$
and $\delta$ by using (\ref{eIIIxxix}) leads to
\begin{eqnarray}\label{eAxvi}
\sum_{a_1} f^{M^3_0}_{1\bar{a}_1a_12}
(\th_2-i\pi,\th_1-i\pi,\th_1-i\eta,\th_2-i\delta)
&=& \cr\cr
&& \hskip -3.in  
-{\Delta\pi^5 \over 8} {16\over \pi^4} ({\cosh(\th_1)\over i\delta}
+{\cosh(\th_2)\over i\eta})
(\prod \psi \big|_{\delta=0,\eta=0} )
G^{m_3}_{1\bar{a}_1a_12}(\th_2-i\pi,\th_1-i\pi,\th_1,\th_2)\cr\cr
&& \hskip -3in -{\Delta\pi^5 \over 8} {16\over \pi^4} 
\bigg\{ \cosh (\th_1) \del_{-i\delta} \bigg( (\prod \psi)
G^{m_3}_{1\bar{a}_1a_12}(\th_2-i\pi,\th_1-i\pi,\th_1-i\eta,\th_2-i\delta)
\bigg)\big|_{\eta=0,\delta=0} \cr\cr
&& \hskip -2.75in
+ \cosh (\th_2) \del_{-i\eta} \bigg( (\prod\psi)
G^{m_3}_{1\bar{a}_1a_12}
(\th_2-i\pi,\th_1-i\pi,\th_1-i\eta,\th_2-i\delta)\bigg)
\big|_{\eta=0,\delta=0} 
\bigg\},
\end{eqnarray}
where $\prod\psi$ is given in this case by
$$
\prod \psi = \psi(\th_{21})
\psi (\th_{21}-i\pi+i\eta)\psi(\th_{12}-i\pi+i\delta)
\psi(\th_{12}-i\eta+i\delta).
$$
Discarding the pole terms (the first set of terms on the r.h.s. of
(\ref{eAxvi})) and evaluating the remainder leaves us with
the desired connected form factor
\begin{equation}\label{eAxvia}
\sum_{a_1} f^{M^3_0}_{1\bar{a}_1a_12}
(\th_2-i\pi,\th_1-i\pi,\th_1,\th_2)_{\rm c} = 
{i2\pi\Delta} {6\pi^2\cosh(\th_1) + (5\pi^2+2\th_{12}^2)\cosh(\th_2)
\over (4\pi^2+\th_{12}^2)(\pi^2+\th_{12}^2)}.
\end{equation}
Combining (\ref{eAxvia}) and (\ref{eAvi}) with (\ref{eAxiv}) we find
\begin{eqnarray}\label{eAxvii}
C_{24}(\om = 0,k=0) &=& {\beta\Delta\over 4\pi}
\int d\th_1d\th_2 e^{-\beta\Delta(\cosh(\th_1)+\cosh(\th_2))}\cr
&& \hskip .25in \times
{6\pi^2\cosh(\th_1) + (5\pi^2+2\th_{12}^2)\cosh(\th_2)
\over (4\pi^2+\th_{12}^2)(\pi^2+\th_{12}^2)}
+ {\rm ~three~other~terms;}\cr\cr
&=& {11\Delta\beta\over 4\pi^3} K_0(\beta\Delta)K_1(\beta\Delta)
+ {\cal O}({T\over\Delta}e^{-\beta\Delta}) + {\rm three~other~terms}.
\end{eqnarray}
To arrive at the last line we have dropped terms polynomial in $\th_{12}$.
This leads to errors of ${\cal O}({T\over\Delta}e^{-\beta\Delta})$.
The remaining three terms make equal contributions to $C_{24}$.  We
thus finally have
\begin{eqnarray}\label{eAxviii}
C_{24}(\om = 0,k=0) &=& {\beta\Delta\over \pi}
\int d\th_1d\th_2 e^{-\beta\Delta(\cosh(\th_1)+\cosh(\th_2))}\cr
&& \hskip 1.25in \times{6\pi^2\cosh(\th_1) + (5\pi^2+2\th_{12}^2)\cosh(\th_2)
\over (4\pi^2+\th_{12}^2)(\pi^2+\th_{12}^2)}.
\end{eqnarray}
We note that in regulating the form factor for $C_{24}$ we do not
allow the infinitesimal imaginary pieces to affect the spatial dependence
of the form factor, i.e. we do not write
$$
f^{M^3_0}_{1\bar{a}_1a_12}
(\th_2-i\pi,\th_1-i\pi,\th_1+i\eta,\th_2+i\delta,x)
= e^{i\Delta x (i\delta \cosh(\th_2) + i\eta \cosh (\th_1))}
(\cdots)
$$
If we were to do so we would find an additional
term coming from expanding $\exp (i\Delta x \cdots)$ in $\eta$
and $\delta$.  
However generically such terms lead to a violation of translation
invariance and as such should not be included.  We moreover know that
such terms would violate the equivalence of the form factor computation
with the expression for the susceptibility coming the thermodynamic
Bethe ansatz.

We go through a similar procedure with $C_{25}$ and find 
an identical result:
$C_{25}(\om=0,k=0) = C_{24}(\om=0,k=0)$.
That we do so is significant.  We might have approached the calculation
equally validly
by ordering the in and out states
in the thermal trace and resolution of identity such that
$\th_1 < \th_2$ and $\th_4 < \th_3$ (and correspondingly multiplying the
expressions in (\ref{eAxi}) by 4).  If we had done so
we would find that in this case $C_{25} = 0$ and $C_{24}$
is twice its current value.  Of course both approaches must yield
the same answer.  But to do so we need $C_{25}(\om=0,k=0) = C_{24}(\om=0,k=0)$.
Given the regularization of the form factors one must do to compute
$C_{25}$, it is not a priori that this will be the case.  That it
is is a non-trivial check of our regularization procedure.

The final term we must evaluate is $C_{26}$.
Fourier transforming as before we find
\begin{eqnarray}\label{eAxix}
C_{26} (\omega = 0 ,k =0)
&=& {1\over 4\Delta^2} \int \dto \dtt \dttr \dtf \cr\cr
&& \times 2\pi 
\delta(\sinh(\th_3)+\sinh(\th_4)-\sinh(\th_2)-\sinh(\th_1))\cr\cr
&& {e^{-\beta\Delta(\cosh(\th_3)+\cosh(\th_4))}
(1-e^{-\beta\Delta(\cosh(\th_3)+\cosh(\th_4)-\cosh(\th_1)-\cosh(\th_2))})
\over \cosh(\th_1)+\cosh(\th_2)-\cosh(\th_3)-\cosh(\th_4)}\cr\cr
&&\times \sum_{a_1a_2a_3a_4}
f^{M^3_0}_{\bar{a}_3\bar{a}_4a_1a_2}
(\th_3-i\pi,\th_4-i\pi,\th_1,\th_2)_{\rm c}\cr
&& \hskip 1in \times 
f^{M^3_0}_{\bar{a}_2\bar{a}_1a_4a_3}
(\th_2-i\pi,\th_1-i\pi,\th_4,\th_3)_{\rm c}.
\end{eqnarray}
As the 4-particle form factors are proportional to 
$(\sinh(\th_3)+\sinh(\th_4)-\sinh(\th_2)-\sinh(\th_1))$ (the Lorentz
pre-factor for the matrix element)
one might believe it is immediate that this expression vanishes 
once the Fourier transform, $\lim_{k\rightarrow 0}\int dx e^{ikx}$,
is taken and so makes
no contribution to the susceptibility.  However the need
to regulate the form factor leaves this ambiguous.  Nevertheless, after
the regulation $C_{26}$ ends up making no contribution to the 
susceptibility.  It will however make a contribution to the NMR
relaxation rate.  Hence some of the details needed to compute $C_{26}$
will be dealt with in the context of that computation (see Appendix B
and Section 2.C).

\vfill\eject

\newcommand{\corb}{{\langle M^1_0(0,t ) M^1_0 (0,0) \rangle}}
\newcommand{\mmb}{{M^1_0(0,t ) M^1_0 (0,0)}}
\newcommand{\mob}{{M^1_0(0,t )}}
\newcommand{\mtb}{{M^1_0(0,0)}}
\newcommand{\ff}{{f^{M^1_0}_{23}}}

\section{Computation of the Correlator for the NMR Relaxation Rate, $\tn$}

In order to compute $\tn$ we must evaluate the correlator,
\begin{equation}\label{eBi}
C(x=0,\om = 0) = \int dt e^{i\om t} \corb .
\end{equation}
The lowest order contribution 
arising from the evaluation of the thermal trace
takes the form
\begin{eqnarray}\label{eBii}
\mmb_{\rm lowest~order}\equiv C_1(t) &=& \int \dt \dto \ebt \sum_{aa_1}
e^{-it\Delta(\cosh(\th_1)-\cosh (\th))+it(H_{s_{a_1}}-H_{s_a})}\cr\cr
&& \hskip -1in 
\times e^{\beta H s_a} \lb A_a(\th ) | \mtb |A_{a_1}(\th_1)\rb
\lb A_{a_1}(\th_1)| \mtb |A_a(\th)\rb,
\end{eqnarray}
where $S_a$ is the spin of particle $a$.  
We have assumed the field, H, is aligned
along the 3-direction.
Although we perform the calculation at finite H, the form-factors themselves
retain their $H=0$ form, a feature of the model's underlying
integrability.  Finite H merely breaks the degeneracy of the triplet
state with the consequent energy shifts seen above.  For the purposes of
this computation, we are interested in the regime $H \ll T \ll \Delta$.
This permits setting $e^{\beta H s_a}$ to 1, provided we are willing
to tolerate errors of $\CO (H/T)$.
Performing then the sums, $\sum_{aa_1}$, over the different types
of excitations leaves us with
\begin{eqnarray}\label{eBiii}
C_1 (t) &=& - 2\int \dt \dto \ebt e^{-it\Delta(\cosh(\th_1)-\cosh (\th))}
\cos (Ht) \cr\cr
&&\hskip .5in \times 
\ff (\th-i\pi,\th_1)\ff(\th_1-i\pi,\th )(1+\CO (H/\Delta)).
\end{eqnarray}
Substituting the expression for the 
form-factors, $\ff$, from Section 3 into the above, and then
performing the necessary
Fourier transform, leaves us with
\begin{equation}\label{eBiv}
C_1(\om =0) = {2\Delta\over\pi}\int d\th 
{\ebt \cosh^2(\th) \over \sqrt{\sinh^2(\th ) +{2H\over\Delta}\cosh (\th )}}
(1+\CO (H/\Delta) +\CO (H/T)).
\end{equation}
For $T\ll \Delta$ this reduces to 
$C_1(\om =0 ) \approx {2\Delta\over\pi}e^{-\bd} (\log (4T/H)-\gamma )$,
where $\gamma$ is Euler's constant.  This is the
result found in \cite{sagi} - a logarithmic dependence on $H$ indicative
of ballistic transport.

The next order in the computation, essentially computing terms of 
$\CO (e^{-2\bd})$, is of the form
\begin{eqnarray}\label{eBv}
C_2 (t) \equiv {1\over 4} \sum_{a_1a_2a_3a_4} \int \dto \dtt \dttr \dtf 
\lb A_{a_1}(\th_1)A_{a_2}(\th_2) | \mob | A_{a_3}(\th_3 ) A_{a_4}(\th_4 )\rb \cr
&& \hskip -2.5in \times
\lb A_{a_4}(\th_4)A_{a_3}(\th_3) | \mtb | A_{a_2}(\th_2 ) A_{a_1}(\th_1 )\rb
\cr\cr
&& \hskip -4.5in \equiv C_{21}(t) + C_{22}(t) + C_{23}(t) + C_{24}(t) + C_{25}(t) 
+ C_{26}(t).
\end{eqnarray}
Here we have introduced the same notation employed to evaluate the second
order contribution to the susceptibility.  The definitions
of $C_{2i}$ are the same as those in (\ref{eAxi}) but for
changing $M^3_0$ to $M^1_0$ and shifting energies by a Zeeman term.
As in the susceptibility computation, $C_{21}(t)$
is an ill-defined term proportional to $\delta (0)$, but is cancelled
off by similar terms coming from the partition function.  Similarly, $C_{22}$
and $C_{23}$ are disconnected terms related to $C_1$.  They give
a contribution of the form
\begin{eqnarray}\label{eBvi}
C_{22} (\om =0) + C_{23} (\om =0) &=&
{2\Delta\over\pi}\int d\th 
{\ebt \cosh^2(\th) \over \sqrt{\sinh^2(\th ) +{2H\over\Delta}\cosh (\th )}}
\times (-3e^{-\bd\cosh (\th )});\cr\cr
&=& -{6\over \pi}\Delta e^{-2\bd}\sqrt{2\pi\over \bd}(\log ({2T\over H})-\gamma).
\end{eqnarray}
If we were to add similarly disconnected terms coming from matrix elements
with a greater number of particle numbers, we would find a resummation
of the form:
\begin{eqnarray}\label{eBvii}
C_{21} (\om =0) +
C_{22} (\om =0) + C_{23} (\om =0) + {\rm higher~order~disconnected~terms}
&& \cr\cr
&& \hskip -3.5in = {2\Delta\over\pi}\int d\th 
{\ebt \cosh^2(\th) \over \sqrt{\sinh^2(\th ) +{2H\over\Delta}\cosh (\th )}}
{1 \over 1 + 3e^{-\bd\cosh (\th )}}.
\end{eqnarray}
This type of resummation was discussed in Section 2.A.2.

The remaining terms are connected.  $C_{24}(t)$ is given by
(we again set terms of the form $e^{\pm\beta H}$ to 1)
\begin{eqnarray}\label{eBviii}
C_{24}(t) = {1 \over 4}\sum_{a_1a_2a_3} \int \dto \dtt \dttr 
e^{-\beta\Delta(\cosh(\th_1)+\cosh(\th_2))}
e^{-it\Delta(\cosh(\th_3)-\cosh(\th_2))}
\cos (Ht) \cr\cr
&& \hskip -4.2in \times 
\Bigg\{ f^{M^1_0}_{\bar{a}_3,\bar{a}_1,a_1,a_2}
(\th_3-i\pi,\th_1-i\pi,\th_1,\th_2)_{\rm c}
f^{M^1_0}_{\bar{a}_2,a_3}(\th_2-i\pi,\th_3)\cr
&& \hskip -5.2in +
\sum_{a_4a'_1a'_2a'_4} S^{a'_4a'_2}_{a_4a_3}(\th_{21}) 
S^{a'_2a'_1}_{a_2a_1}(\th_{13})
f^{M^1_0}_{\bar{a}_2,\bar{a}_1,a_4,a_3}
(\th_1-i\pi,\th_3-i\pi,\th_2,\th_1)_{\rm c}
f^{M^1_0}_{\bar{a}'_4,a'_1}(\th_2-i\pi,\th_3)\cr
&& \hskip -2.5in +~(\th_2 \leftrightarrow \th_3)\Bigg\}.
\end{eqnarray}
To evaluate this expression we must again regulate the four particle
form-factors appearing in the above by removing the singularities arising
when two rapidities equal one another.  For example, we regulate
the first four particle form-factor appearing in the above via
\begin{eqnarray}\label{eBix}
f^{M^1_0}_{\bar{a}_3,\bar{a}_1,a_1,a_2}
(\th_3-i\pi,\th_1-i\pi,\th_1,\th_2)_{\rm c} 
&=& \cr\cr
&& \hskip -2.10in 
{\pi^3\over 2}\psi (\th_{32}-i\pi)\bigg\{ \cosh (\th_1)
(\prod\psi) G^{M^1_0}_{\bar{a}_3,\bar{a}_1,a_1,a_2}
(\th_3-i\pi,\th_1-i\pi,\th_1,\th_2) \cr\cr
&& \hskip -2.10in +(\sinh(\th_2)-\sinh(\th_3))
\del_{-i\eta} \bigg((\prod\psi) G^{M^1_0}_{\bar{a}_3,\bar{a}_1,a_1,a_2}
(\th_3-i\pi,\th_1-i\pi,\th_1-i\eta,\th_2)\bigg)\bigg\},
\end{eqnarray}
where 
$\prod\psi = \psi (\th_{31}) \psi (\th_{31}-i\pi+i\eta) 
\psi (\th_{12}-i\pi)\psi(\th_{12}-i\eta)$.
Regulating the other form-factors similarly, we find after
a long computation
\begin{eqnarray}\label{eBx}
C_{24}(\om =0) &=& {\Delta\pi\over 256}
\int d\th_1d\th_2 e^{-\bd(\cosh(\th_1)+\cosh(\th_2))}
\Bigg\{ \bigg( {1\over |\sinh(\th_3)|}(\sinh(\th_2)-\sinh(\th_3))\cr\cr
&& \hskip -.75in \times 
\bigg[{\th_{23}\coth^2(\th_{23}/2)\over (\th_{23}^2+\pi^2)})\bigg]^2
\bigg\{ 12\thinspace\th_{23}\cosh(\th_1)+12(\sinh(\th_2)-\sinh(\th_3))
({\th_{13}\over\th_{12}}+{\th_{12}\over\th_{13}}-{1\over6})\bigg\}\cr\cr
&& \hskip -.75in \times (1+\CO(\th_{23})^2+\CO(\th_{12})^2+\CO(\th_{13})^2))
\bigg)\bigg|_{\th_3=\cosh^{-1}(\cosh(\th_2)+H)} 
+ (H \leftrightarrow -H) \Bigg\} \cr\cr
&=& {17 \Delta \over 2\pi^3}e^{-2\bd}\sqrt{2\pi\over\bd} 
\big(\log ({4T\over H})-\gamma\big)
(1+\CO (H/T)+ \CO(T/\Delta)).
\end{eqnarray}
We perform a similar procedure on $C_{25}$.  As with the susceptibility,
$C_{25}$ must and does generate an identical contribution to $C_{24}$.

The remaining term to evaluate is $C_{26}$.  This term made no contribution
to the susceptibility but does make a contribution to the relaxation rate,
$\tn$.
$C_{26}$ takes the form
\begin{eqnarray}\label{eBxi}
C_{26}(t) = {1\over 4}\sum_{a_1a_2a_3a_4} \int \dto \dtt \dttr \dtf
e^{-\beta\Delta(\cosh(\th_1)+\cosh(\th_2))}\cr\cr\cr
&& \hskip -3.25in 
\times 
~~e^{-it\Delta(\cosh(\th_3)+\cosh(\th_4)-\cosh(\th_1)-\cosh(\th_2))} 
e^{itH(S_{a_3}+S_{a_4}-S_{a_1}-S_{a_2})}\cr\cr
&& \hskip -3.25in \times 
~~f^{M^1_0}_{\bar{a}_2,\bar{a}_1,a_4,a_3}
(\th_2-i\pi+i\ep_2,\th_1-i\pi+i\ep_1,\th_4,\th_3)_{\rm c} \cr\cr
&& \hskip -3.25in \times 
~~f^{M^1_0}_{\bar{a}_3,\bar{a}_4,a_1,a_2}
(\th_3-i\pi+i\ep_3,\th_4-i\pi+i\ep_4,\th_1,\th_2)_{\rm c}.
\end{eqnarray}
Again we must regulate this expression by discarding terms proportional
to the $1/\ep$'s.  To exhibit such terms we deform the contours $\th_3$ and
$\th_4$ via
$$
\th_3 \rightarrow \th_3 + i\pi;
$$
$$
\th_4 \rightarrow \th_4 + i\pi.
$$
In doing so, we deform through a series of poles whose
residues we thus pick up.  Taking these into account, we end up with
\begin{eqnarray}\label{eBxii}
C_{26} (t) &=& -{i\over 8\pi^5}\sum_{a_1a_2a_3a_4}
\int d\th_1d\th_2d\th_4 \ebtot e^{-i\Delta t\cosh(\th_4)}\cos (Ht)\cr\cr
&& \times \Bigg\{ e^{i\Delta t\cosh(\th_2)}
{f^{M^1_0}_{\bar{a}_2,\bar{a}_1,a_4,a_3}
(\th_2-i\pi+i\ep_2,\th_1-i\pi+i\ep_1,\th_4,\th_3)\over
\psi(\th_{13}-i\pi+i\ep_1)}\bigg|_{\th_3=\th_1} \cr\cr
&& \hskip .5in \times f^{M^1_0}_{\bar{a}_3,\bar{a}_4,a_1,a_2}
(\th_1-i\pi+i\ep_3,\th_4-i\pi+i\ep_4,\th_1,\th_2)\cr\cr
&&  \hskip .3in + 
e^{i\Delta t\cosh(\th_1)}{f^{M^1_0}_{\bar{a}_2,\bar{a}_1,a_4,a_3}
(\th_2-i\pi+i\ep_2,\th_1-i\pi+i\ep_1,\th_4,\th_3)\over
\psi(\th_{23}-i\pi+i\ep_2)}\bigg|_{\th_3=\th_2} \cr\cr
&& \hskip .5in \times f^{M^1_0}_{\bar{a}_3,\bar{a}_4,a_1,a_2}
(\th_2-i\pi+i\ep_3,\th_4-i\pi+i\ep_4,\th_1,\th_2) \Bigg\}\cr\cr
&-& {i\over 8\pi^5}\sum_{a_1a_2a_3a_4}
\int d\th_1d\th_2d\th_3 \ebtot e^{it\Delta\cosh(\th_3)}\cos (Ht)\cr\cr
&& \times \Bigg\{ e^{i\Delta t\cosh(\th_2)}
{f^{M^1_0}_{\bar{a}_2,\bar{a}_1,a_4,a_3}
(\th_2-i\pi+i\ep_2,\th_1-i\pi+i\ep_1,\th_4,\th_3+i\pi-i\ep_3)\over
\psi(\th_{14}-i\pi+i\ep_1)}\bigg|_{\th_4=\th_1}\cr\cr
&& \hskip .5in \times f^{M^1_0}_{\bar{a}_3,\bar{a}_4,a_1,a_2}
(\th_3,\th_1-i\pi+i\ep_4,\th_1,\th_2)\cr\cr
&&  + e^{it\Delta\cosh (\th_1)}
{f^{M^1_0}_{\bar{a}_2,\bar{a}_1,a_4,a_3}
(\th_2-i\pi+i\ep_2,\th_1-i\pi+i\ep_1,\th_4,\th_3+i\pi-i\ep_3)\over
\psi(\th_{24}-i\pi+i\ep_2)}\bigg|_{\th_4=\th_2} \cr\cr
&& \hskip .5in \times f^{M^1_0}_{\bar{a}_3,\bar{a}_4,a_1,a_2}
(\th_3,\th_2-i\pi+i\ep_4,\th_1,\th_2) \Bigg\}\cr\cr
&+& {1\over 64\pi^4}\sum_{a_1a_2a_3a_4}\int d\th_1d\th_2d\th_3d\th_4
\ebtot \cr\cr
&& 
\hskip .4in
\times e^{it\Delta(\cosh (\th_1) +\cosh (\th_2) +\cosh(\th_3) + \cosh(\th_4))}
\cos (Ht)
\cr\cr
&& \hskip .5in \times
f^{M^1_0}_{\bar{a}_2,\bar{a}_1,a_4,a_3}
(\th_2-i\pi,\th_1-i\pi,\th_4+i\pi,\th_3+i\pi)
f^{M^1_0}_{\bar{a}_3,\bar{a}_4,a_1,a_2}
(\th_3,\th_4,\th_1,\th_2)\cr\cr
&\equiv& \sum_{i=1}^5 C_{26i}(t).
\end{eqnarray}
As we are interested in $C_{26}(\om \sim 0)$, we immediately see
that the last three terms, $C_{263}$ to $C_{265}$, may be neglected
as they are only non-zero
for frequencies, $\omega$, in excess of $2\Delta$ (provided $H \ll \Delta$).

Focusing then upon $C_{261}$, we obtain upon performing the necessary regulation
\begin{eqnarray}\label{eBxiii}
C_{261}(t) &=& i{\Delta^2\pi^3\over 128}\int d\th_1d\th_2d\th_4
\ebtot \cos (Ht) e^{-i\Delta t(\cosh(\th_4)-\cosh(\th_2))} \cr\cr
&& \times (\sinh (\th_4)-\sinh(\th_2))(\prod_1\psi )
G^{m^1_0}_{\bar{a}_2,\bar{a}_1,a_4,a_3}
(\th_2-i\pi,\th_1-i\pi,\th_4,\th_1) \cr\cr
&& \times \bigg\{ \cosh(\th_1)(\prod_2\psi )
G^{m^1_0}_{\bar{a}_3,\bar{a}_4,a_1,a_2}
(\th_1-i\pi,\th_4-i\pi,\th_1,\th_2)\nonumber \\
&& \hskip .2in + (\sinh (\th_2)-\sinh( \th_4))
\del_{-i\ep}
\bigg( (\prod_2\psi )
G^{m^1_0}_{\bar{a}_3,\bar{a}_4,a_1,a_2}
(\th_1-i\pi+i\ep,\th_4-i\pi,\th_1,\th_2)\bigg)\cr\cr
\prod_1 &=& \psi(\th_{21})\psi(\th_{24}-i\pi)\psi(\th_{21}-i\pi)
\psi(\th_{14}-i\pi)\psi(\th_{41})\cr\cr
\prod_2 &=& \psi(\th_{14}+i\ep)\psi(\th_{12}-i\pi+i\ep)\psi(\th_{41}-i\pi)
\psi(\th_{42}-i\pi)\psi(\th_{12})
\end{eqnarray}
To evaluate the above expression, we first Fourier transform which then
leads us to consider the following expressions:
\begin{eqnarray}\label{eBxiv}
&& (\prod_1\psi)(\prod_2\psi) 
G^{m^1_0}_{\bar{a}_2,\bar{a}_1,a_4,a_3}G^{m^1_0}_{\bar{a}_3,\bar{a}_4,a_1,a_2}
- (\th_i \leftrightarrow -\th_i) = 0\cr\cr
&& (\prod_1\psi)
G^{m^1_0}_{\bar{a}_2,\bar{a}_1,a_4,a_3}
\del_{-i\ep} \bigg( (\prod_2\psi) G^{m^1_0}_{\bar{a}_3,\bar{a}_4,a_1,a_2}\bigg)
+ (\th_i\leftrightarrow -\th_i) \nonumber\\
&& \hskip .5in = (\prod_1\psi)
G^{m^1_0}_{\bar{a}_2,\bar{a}_1,a_4,a_3}G^{m^1_0}_{\bar{a}_3,\bar{a}_4,a_1,a_2}
\bigg(\del_{-i\ep}(\prod_2\psi) + (\th_i \leftrightarrow -\th_i)\bigg) \nonumber\\
&& \hskip .5in = i12\pi \bigg[{\th_{24}^2\coth^4(\th_{24}/2)\over (\th_{24}^2+\pi^2)}\bigg]^2(1+\CO (\th_{14}^2)+\CO (\th_{12}^2)+\CO (\th_{24}^2)).
\end{eqnarray}
Putting everything together then yields
\begin{eqnarray}\label{eBxv}
C_{261}(\om =0) &=& {\Delta3\pi^5\over 16}
\int d\th_1d\th_2 e^{-\bd(\cosh(\th_1)+\cosh(\th_2))}
\Bigg\{ \bigg( {1\over |\sinh(\th_4)|}(\sinh(\th_2)-\sinh(\th_4))^2\cr\cr
&& \hskip -.75in \times 
\bigg[{\th_{24}^2\coth^4(\th_{24}/2)\over (\th_{24}^2+\pi^2)}\bigg]^2\nonumber \\
&& \hskip -.75in \times (1+\CO(\th_{24})^2+\CO(\th_{12})^2+\CO(\th_{14})^2))
\bigg)\bigg|_{\th_4=\cosh^{-1}(\cosh(\th_2)+H)} 
+ (H \leftrightarrow -H) \Bigg\} \cr\cr
&=& 12 \pi \Delta e^{-2\bd}\sqrt{2\pi\over\bd}\big( \log ({4T\over H})-\gamma\big)
(1+\CO (H/T)+ \CO(T/\Delta)).
\end{eqnarray}
The evaluation of $C_{262}$ yields an identical contribution
to the relaxation time.

\begin{figure}
\centerline{\hskip -1in\psfig{figure=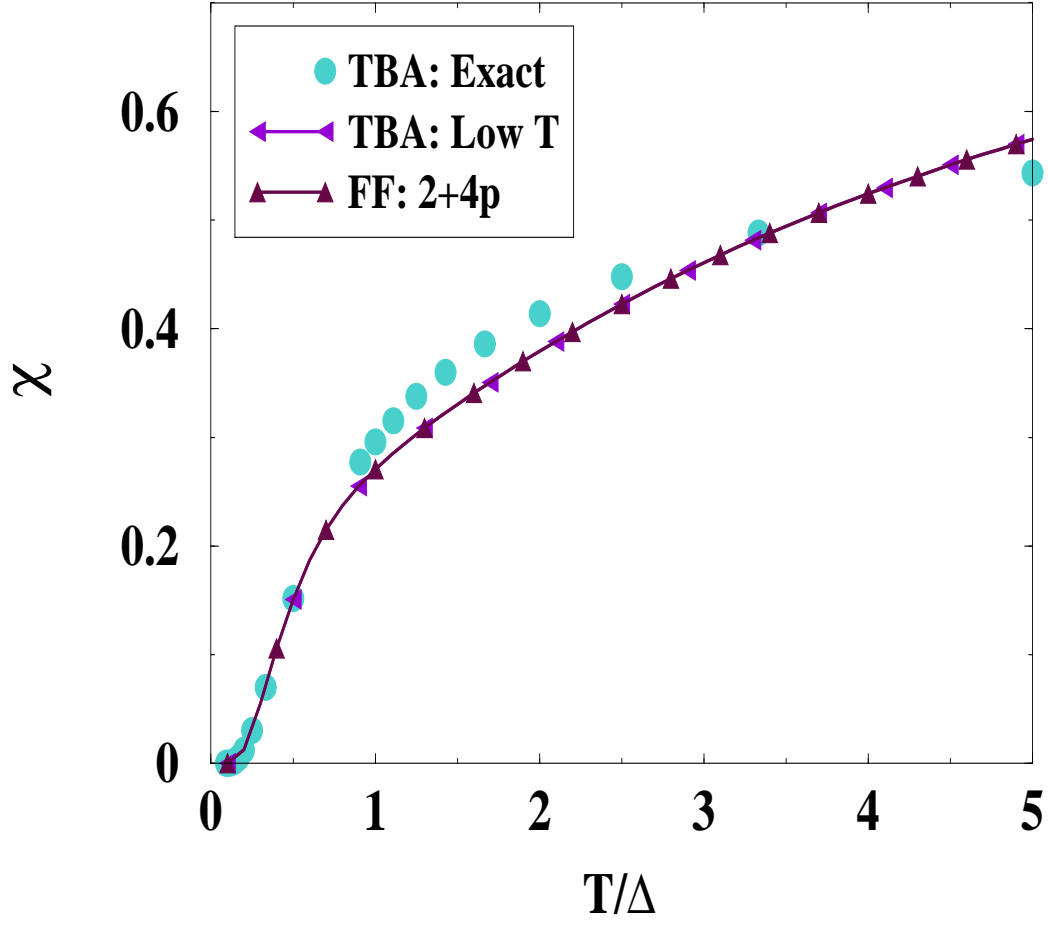,angle=-90,height=5in,width=5in}}
\caption{Plots of the zero-field
susceptibility computed both from the TBA equations
and from the form factor expansions.
The first of these is an
exact numerical solution of the TBA equations for the O(3) sigma model.  
The second is arrived from a
small temperature expansion in powers of $e^{-\beta\Delta}$ of these same
equations.  
The final plot gives the form factor computation of the
susceptibility.  We have truncated the form factor expansion
at the four particle level.}
\end{figure}

\vfill\eject

\begin{figure}
\centerline{\hskip -1in\psfig{figure=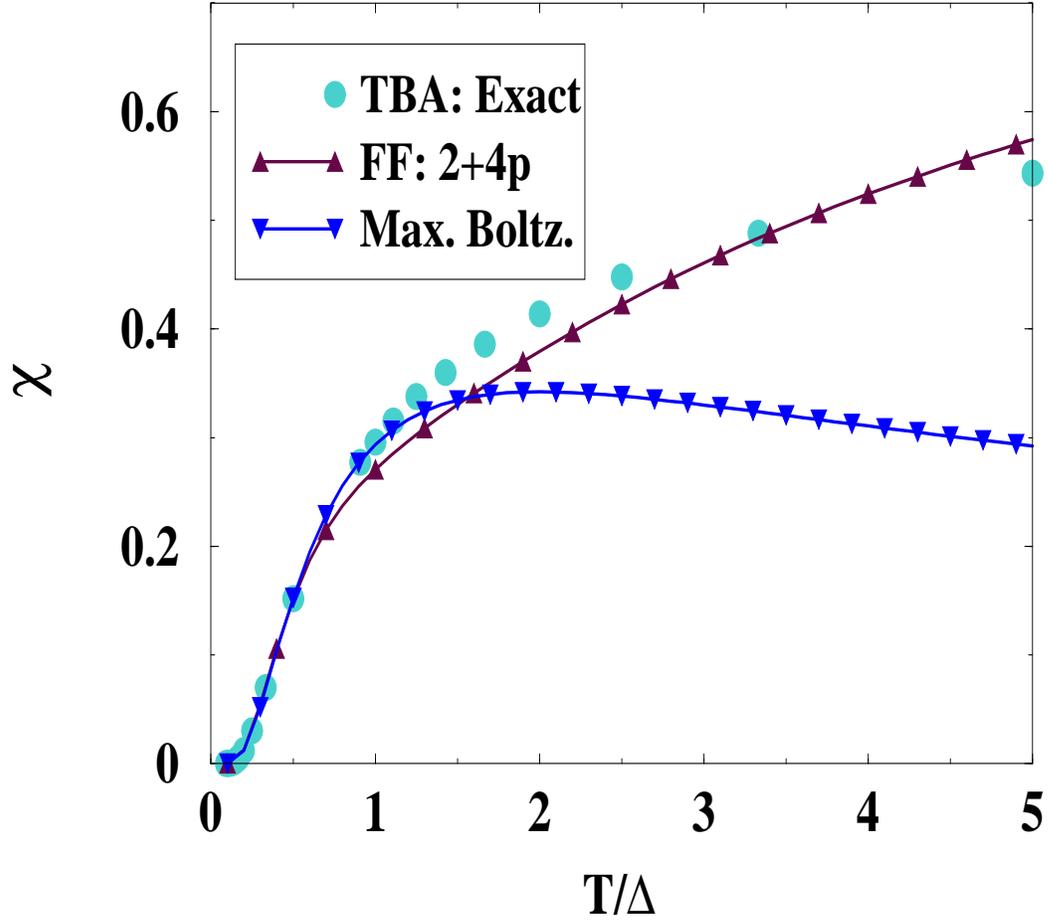,angle=-90,height=5in,width=5in}}
\caption{
The zero-field susceptibility of a Maxwellian gas is compared here
to both the exact susceptibility of the O(3) NLSM and
the susceptibility of the O(3) NLSM computed via a form
factor expansion.}
\end{figure}

\vfill\eject

\begin{figure}
\centerline{\hskip -1in\psfig{figure=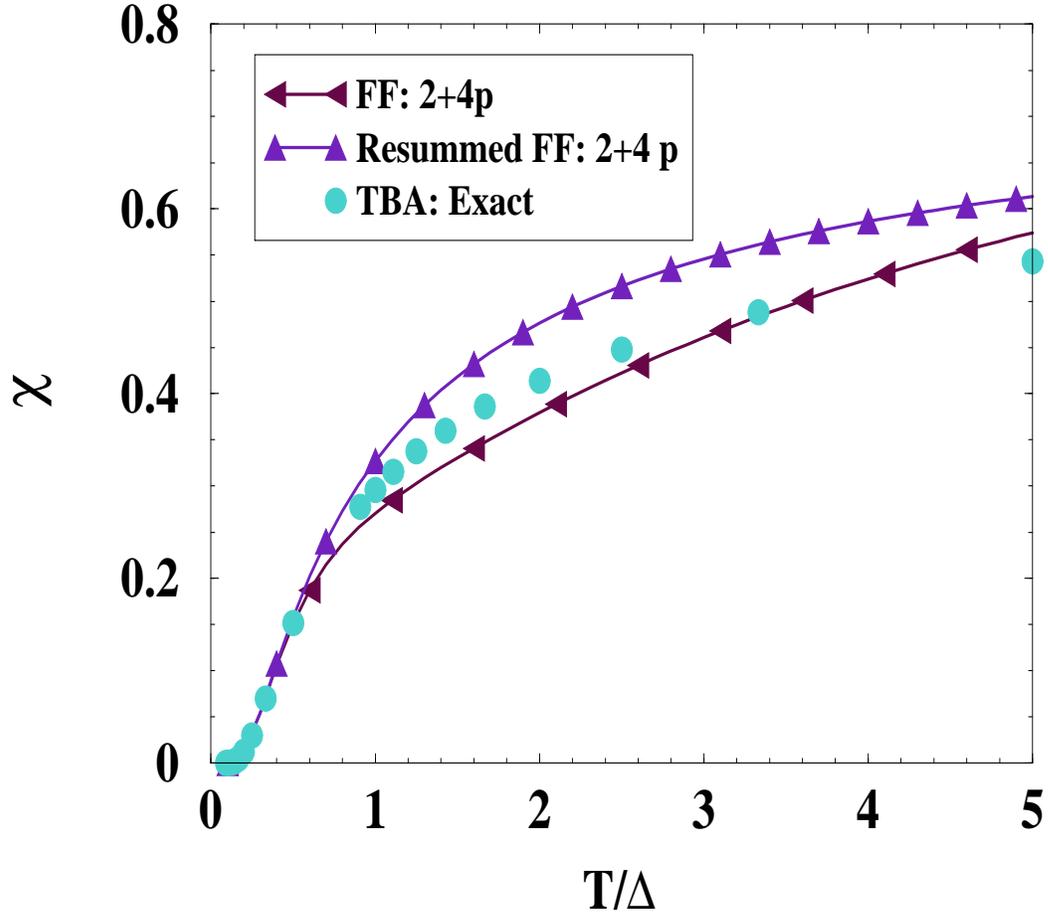,angle=-90,height=5in,width=5in}}
\caption{The zero-field susceptibility of the O(3) NLSM as computed
using a resummed form factor expansion
is compared both 
with the exact result coming from the TBA equations and the unresummed
form factor susceptibility.}

\end{figure}

\vfill\eject

\begin{figure}
\centerline{\hskip -1in\psfig{figure=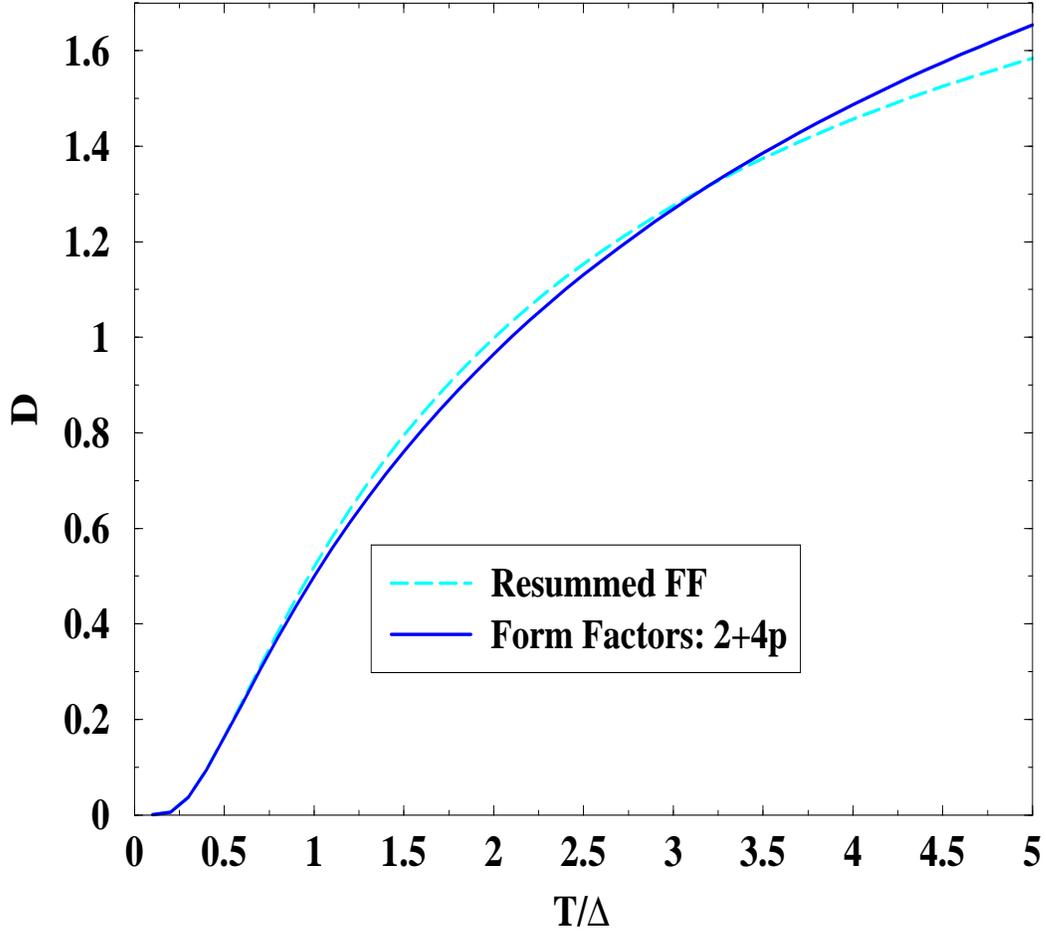,angle=-90,height=5in,width=5in}}
\caption{In this plot we present the form factor
computation of the Drude weight, $D$, of the spin conductance.  
As with the susceptibility,
both the unresummed and resummed computation give roughly the same answer.}

\end{figure}

\vfill\eject

\begin{figure}
\centerline{\hskip -1in\psfig{figure=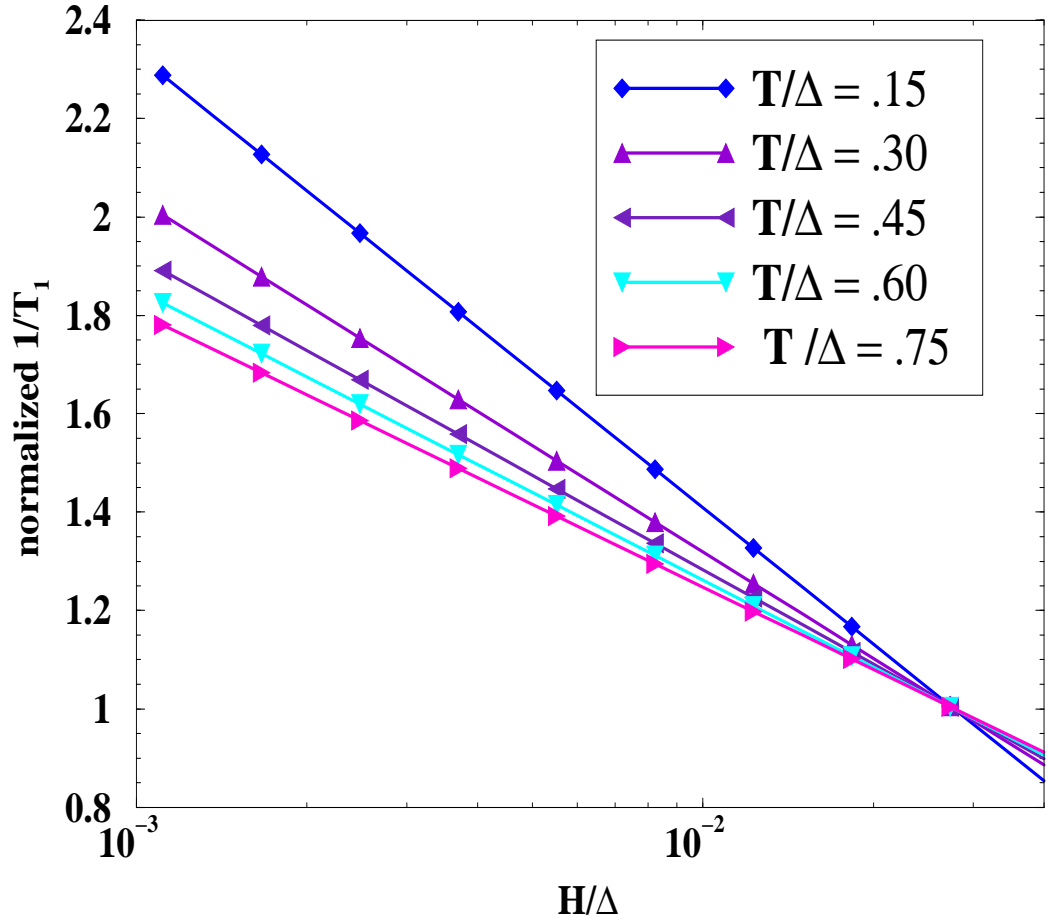,angle=-90,height=5in,width=5in}}
\caption{In this log-linear
plot we present the form factor
computation of the NMR relaxation rate, $1/T_1$, as a function of H
for a variety of temperatures.  We plot a normalized
rate, the ratio of $1/T_1(H)$ with $1/T_1 (H=\Delta/36)$.}
\end{figure}

\end{document}